\documentclass{CUP-JNL-DCE}

\usepackage{latexsym}
\usepackage{graphicx}
\usepackage{makecell}
\usepackage{longtable,tabularx}
\usepackage{booktabs}
\usepackage[version=4]{mhchem}
\usepackage{subfigure}
\usepackage{siunitx}
\usepackage[utf8]{inputenc}
\usepackage{placeins}
\usepackage{multicol,multirow}
\usepackage{amsmath,amssymb,amsfonts}
\usepackage{mathrsfs}
\usepackage{amsthm}
\usepackage{apacite}
\usepackage{rotating}
\usepackage{appendix}
\usepackage[authoryear]{natbib}
\usepackage{ifpdf}
\usepackage[T1]{fontenc}
\usepackage{times}
\usepackage{sourcesanspro}
\usepackage{newtxmath}
\usepackage{textcomp}%
\usepackage{xcolor}%
\usepackage{hyperref}
\usepackage{bm}
\newcommand{\dist}[1]{\left\lVert #1 \right\rVert}

\articletype{RESEARCH ARTICLE}
\jname{Data-Centric Engineering}
\jyear{YEAR}
\jdoi{xx.xx}

\DeclareUnicodeCharacter{03F5}{\ensuremath{\epsilon}}
\begin{document}

\begin{Frontmatter}

\title[Large Wing Model]
{Large Wing Model}

\author*[1]{Howon Lee}\email{hlee981@gatech.edu}\orcid{0000-0001-7942-2318}
\author[1]{Pranay Seshadri}\orcid{0000-0003-1845-3377}
\author[1]{Juergen Rauleder}\orcid{0000-0001-5882-860X}

\authormark{Howon Lee \textit{et al.}}

\address[1]{\orgdiv{Daniel Guggenheim School of Aerospace Engineering}, \orgname{Georgia Institute of Technology}, \orgaddress{\street{270 Ferst Dr, Atlanta}, \state{GA}, \postcode{30332}, \country{USA}}}

\received{\today} 

\abstract{Developing a generalized aerodynamics prediction machine learning model for finite wings with different airfoil sections is challenging due to the vast parameter space and a relative scarcity of available data. This paper presents the Large Wing Model (LWM), a probabilistic machine learning model designed to predict pressure coefficient ($C_p$) distributions using a small, strictly experimental data set. From its uncertainty-aware $C_p$ predictions, the sectional and total wing lift coefficients ($c_l$, $C_L$) and their confidence intervals are calculated. The LWM features a modified deep kernel learning architecture, building a Gaussian Process model in a 15-dimensional space formed by 14 latent variables and the wing spanwise dimension. It is trained on an open-source database of wind tunnel measurements developed for this work. The Bayesian approach ingests uncertainties associated with experimental measurements and data digitization into the model. The model demonstrates satisfactory extrapolation abilities, enabling predictions on wings with new airfoil sections via the physics-driven prior formed from two-dimensional $C_p$ predicted by the Large Airfoil Model. The model accuracy is assessed for three test cases, rectangular wings with varying airfoil sections and operating conditions. For all test cases, the model performed well; the error in $C_L$ did not exceed 1.7\%. The model effectively captures three-dimensional effects such as those induced by wing tip vortices. Furthermore, constraining the posterior predictive space based on known probabilistic descriptions of lift improves the accuracy of the $C_p$ predictions. As a computationally efficient wing $C_p$ prediction model, the LWM facilitates the rapid exploration of the wing design space.}

\begin{policy}[Impact Statement]
This article offers a robust, extrapolatable probabilistic machine learning model designed to predict the pressure distribution over a finite, rectangular wing. The model is exclusively trained on experimental data. Specific contributions of this article include:
\begin{itemize}
\item{a digitized database of historical experimental finite wing pressure measurements,}
\item{utilization of spatially-varying hyperparameters for capturing rapidly developing pressure distortions at the wing tip, such as those induced by wing tip vortices,}
\item{leveraging of physics-driven priors, enabled by probabilistic two-dimensional pressure predictions of a machine learning model previously developed by the authors, and}
\item{a Bayesian framework to condition the posterior predictive distribution on real measurements of linear operators on pressure, such as the lift coefficient.} 
\end{itemize}
\end{policy}
\end{Frontmatter}

\section{Introduction}\label{sec:intro}
\begin{figure}[h!]
\centering
\includegraphics[width=.45\textwidth]{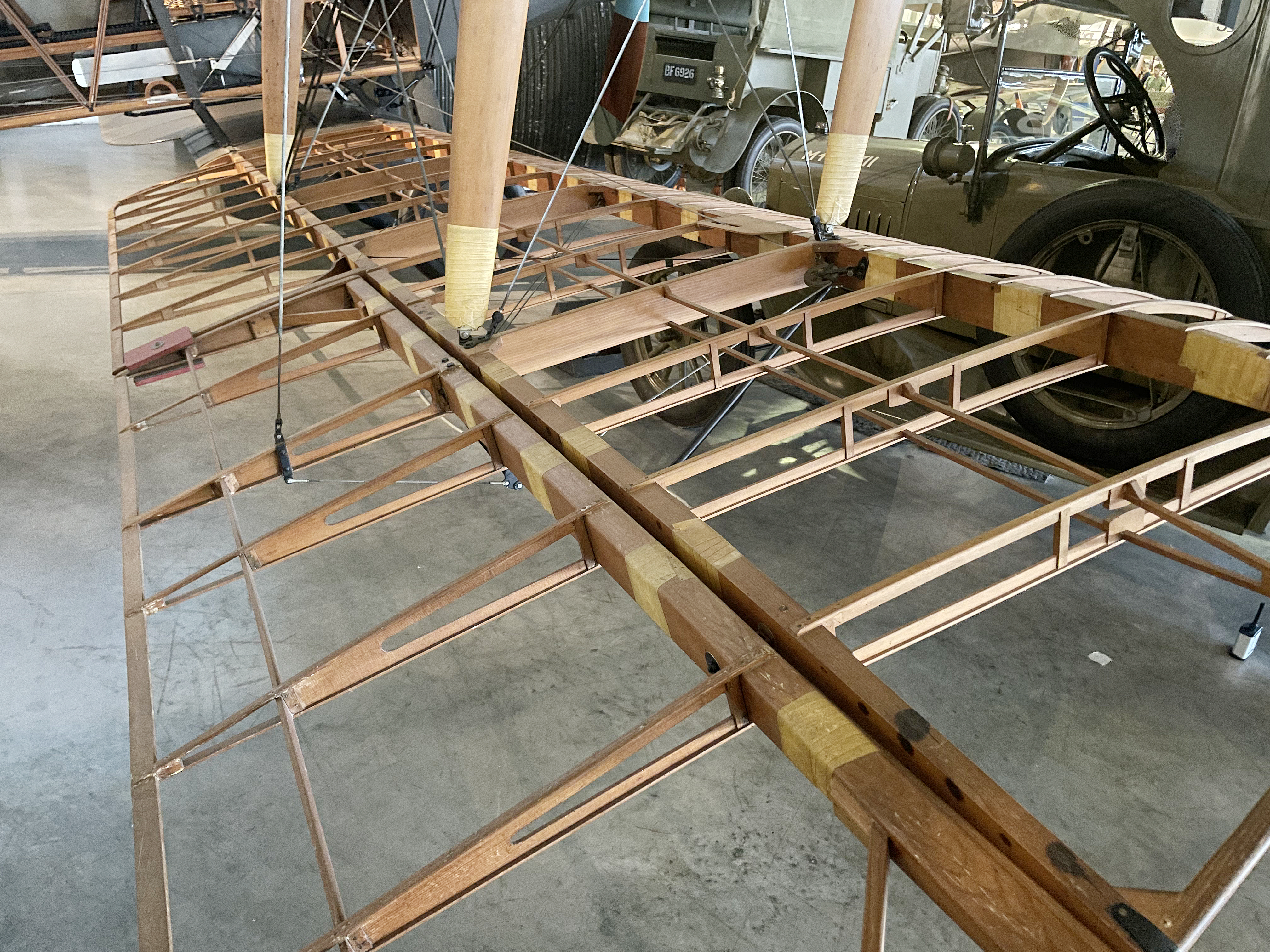}
\caption{Internal structure of the wing of a Bristol F.2 Fighter. The ribs reveal the airfoil profile at various spanwise stations, illustrating how the wing shape is constructed}
\label{fig:wing_photo}
\end{figure}

Aerodynamic force and moment coefficients, such as sectional and total lift coefficients ($c_l$ and $C_L$), are fundamental parameters in aerospace engineering, serving as a cornerstone in the aircraft and wing design and optimization process. As illustrated in Fig.~\ref{fig:wing_photo}, the behavior of a wing is largely driven by its cross-sectional shape, known as an airfoil. For over a century, airfoils have provided the foundational geometric basis for aerodynamic design, with historical airfoil catalogs and wind tunnel measurements continuing to inform both conventional and new configurations. 

Efficient computational methods for calculating the aerodynamic coefficients are critical for effectively navigating the aerodynamic design space. While lower-fidelity methods are commonly employed for this purpose, they often struggle to accurately capture complex aerodynamic phenomena. Although high-fidelity computational fluid dynamics (CFD) simulations can model these effects with improved precision, their computational cost renders them impractical for the initial design phase. Furthermore, such CFD models often require advanced user knowledge to be effective (e.g., to mesh/model the wing tip section or flow transition). This motivates the need for a higher-fidelity pressure coefficient ($C_p$) prediction model that operates at a reduced computational cost, enabling more efficient and informed design space exploration, yet not requiring much advanced user knowledge of the aerodynamic problem. Recent developments in computational power and data-driven modeling have positioned machine learning (ML) as a promising tool for developing such models.

ML methodologies have been highly successful in predicting the aerodynamic properties of two-dimensional airfoils. These efforts include directly predicting the $C_p$~(Yilmaz and German,~\citeyear{yilmaz2017}; Hui et al.,~\citeyear{Hui_2020}; Zhang,~\citeyear{zhang_2023}; Lee et al.,~\citeyear{lee2024_LAM}), or directly predicting the sectional lift, drag, and moment coefficients ($c_l$, $c_d$, and $c_m$) (Zhang et al.,~\citeyear{Zhang_2018}; Liu et al.,~\citeyear{Liu_2022}; Cornelius and Schmitz,~\citeyear{cornelius2024}).

Several studies have extended ML techniques to predict the aerodynamic coefficients of finite wings (Zhao et al.,~\citeyear{zhao2023}; Zhang et al.,~\citeyear{zhang2024}). Most existing studies simplify the problem by fixing either the wing planform geometry or the airfoil section. The simplified approach stems from the significantly larger parameter space of a wing, which can vary in airfoil section, planform geometry, geometric twist, and operating conditions. As a result, a substantially large database is required for accurate analysis, which may be impractical to generate due to its size. Efforts have also been made to develop more generalized models that can be trained on smaller data sets. For instance, Yang et al.~\citeyearpar{yang2024_wing} demonstrated that using distributed geometry inputs improves the prediction accuracy, while reducing the required data set size.

The majority of ML models are dependent on an extensive sweep of CFD simulations for training. However, this reliance introduces several limitations. First, CFD dependence may introduce biases in numerical results, arising from modeling decisions such as the choice of turbulence or transition models. Second, modeling the three-dimensional flow over a wing is inherently challenging, particularly near the wing tip. The accurate representation of wing tip vortices requires high scheme accuracy, high spatio-temporal resolution, and a suitable grid topology (Pereira et al.,~\citeyear{Pereira_wingtip}). Moreover, most turbulence closures based on the Boussinesq Hypothesis often fail to accurately capture wing tip vortex-induced effects when paired with Reynolds-Averaged Navier--Stokes (RANS) solvers~(Wilcox and Chambers,~\citeyear{wingtipvort_issue1}; Bradshaw,~\citeyear{wingtipvort_issue2}; Liu et al.,~\citeyear{LIU2016227}). Lastly, uncertainty quantification for computational data is very nuanced and involved, requiring sophisticated techniques such as field inversion to infer modeling discrepancies in turbulence closures (Singh and Duraisamy,~\citeyear{singh2016}).  

Experimental measurements offer a potential alternative to CFD data as a training data source; however, this requires detailed uncertainty bookkeeping. Measurement uncertainties in wind tunnel experiments can originate from various sources, such as manufacturing tolerances, unsteadiness in the tunnel freestream, as well as the noise characteristics of the measuring equipment. The digitization of publicly available data sets, in particular results that only contain graphs, may also introduce additional uncertainties. Many machine learning architectures, namely neural networks, are deterministic in nature and have difficulty quantifying such uncertainties. 

Adopting a Bayesian approach offers practical techniques to incorporate uncertainty awareness into ML models. For example, Anhichem et al. used Gaussian Process (GP) regression to create surrogate models for pressure distributions over an OAT15A airfoil~(Anhichem et al.,~\citeyear{Anhichem_2024}) and a multi-fidelity model of an RBC12 half-wing-fuselage configuration (Anhichem et al.,~\citeyear{Anhichem_2022}). Another example is the work by Lee et al.~\citeyearpar{lee2024_LAM}, who developed a deep kernel learning model trained on experimental measurements to predict airfoil $C_p$ distributions in an uncertainty-aware manner.  

An essential component of data-driven model development is the quality and scope of the training database. Building a robust model that can reliably generate predictions requires access to a large and comprehensive set of aerodynamic data. Equally important is the public availability of such a database, which facilitates the evaluation of model performance and the comparison of methodologies across the wider community. In recent years, there have been efforts to develop databases and benchmark cases for various aerodynamic applications, including airfoil and wing surface pressure distributions (Bekemeyer et al.,~\citeyear{Bekemeyer2025}; Lee et al.,~\citeyear{lee2024_LAM}; Petrilli et al.,~\citeyear{Petrilli2013}). However, existing repositories of wing pressure data are entirely computational, and thus inherit the limitations previously discussed.

Hence in this paper, the development of the Large Wing Model (LWM) is presented, which includes both a novel machine learning model for predicting wing $C_p$ distributions and an open-source repository of digitized wing experimental data. The predictive ML model employs deep kernel learning, which combines deep neural networks with a Gaussian process model. Its physics-informed prior structure, driven by the Large Airfoil Model (LAM) developed previously by the authors (Lee et al.,~\citeyear{lee2024_LAM}), enables a more universal wing $C_p$ prediction capability from smaller, exclusively experimental training data sets.

The paper is organized as follows. In Section \ref{sec:data}, the digitized database of wind tunnel measurements for finite-wing $C_p$ distributions is presented. This section also details the data mining process and the pre-processing steps used to format the data for the LWM. In Section \ref{sec:model}, an in-depth description of the developed model is provided. The model accuracy is then assessed for different test cases in Section \ref{sec:results}.

\section{Experimental Database}\label{sec:data}
\subsection{Data Mining and Digitization}
\begin{figure}[h!]
\centering
\includegraphics[width=.99\textwidth]{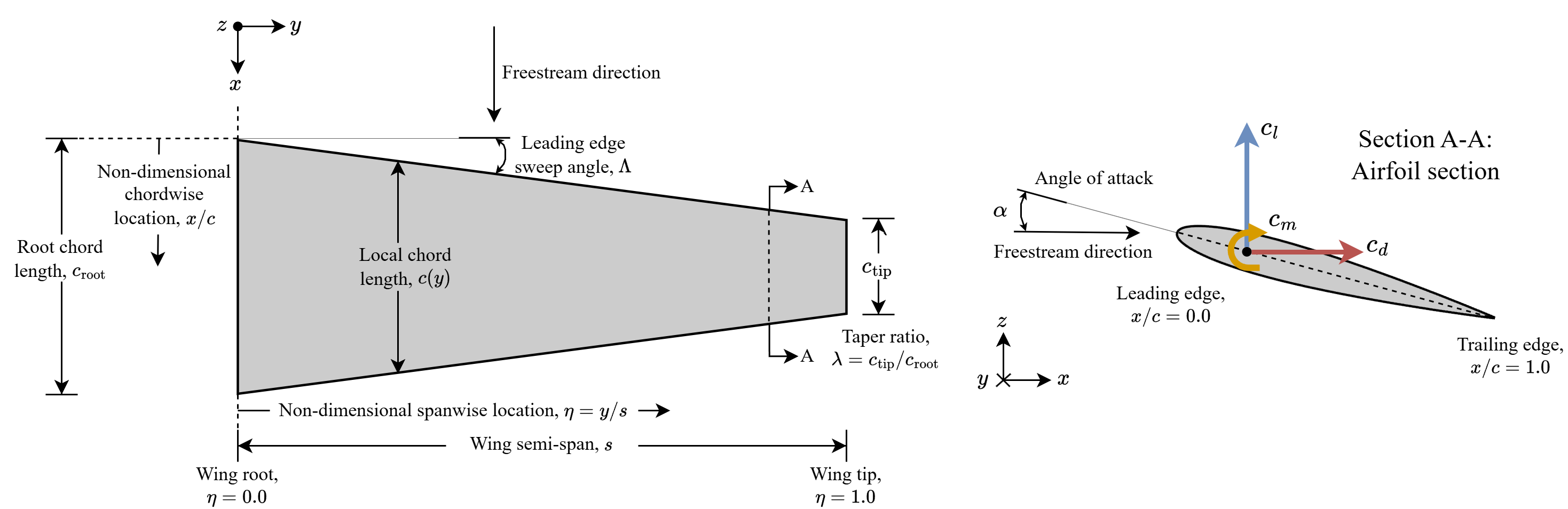}
\caption{Key geometric parameters that are used to define the shape of a finite wing}
\label{fig:wing_geom_definition}
\end{figure}

A data mining campaign was undertaken to create a data set of strictly experimental wing $C_p$ distributions. The digitized database consists pressure measurements from 16 unique wings. These wings vary in airfoil sections and planform geometries, which are defined by the ratio between the semi-span length and root chord length, taper ratio, and leading edge sweep angle ($s/c_\text{root}$, $\lambda$, and $\Lambda$, respectively). These geometric parameters are defined and visualized in Fig.~\ref{fig:wing_geom_definition}.

These wings were tested under 54 operating conditions, defined by a combination of different angles of attack ($\alpha$) and freestream Mach numbers ($M_\infty$). For this work, the data digitization prioritized finite wings in $M_\infty$ up to the transonic regime, under attached flow conditions. The measurements from all operating conditions corresponded to a total of 23,723 individual pressure sensor measurements. 

The sources of experimental data ranges from government reports and technical notes to published research papers. These documents presented their results in either tables or graphs. For tabulated data, the reported values were converted to comma separated values (CSV) files using an online optical character recognition (OCR) tool named ExtractTables~(Saradhi,~\citeyear{extracttable}). For graphical data, the data were digitized using WebPlotDigitizer~(Rohatgi,~\citeyear{webplotdigitizer}), which allows the manual extraction of numerical data from images.

The database is available as an extension to the Airfoil Surface Pressure Information Repository of Experiments (ASPIRE\footnote{Accessible online on \url{https://large-airfoil-model.azurewebsites.net}}). The repository is a large-scale, open-source repository of experimental $C_p$ measurements of airfoils released by Lee et al., \citeyearpar{lee2024_LAM}.

\subsection{Available Data}
As shown in Table~\ref{tab:database_comparison}, compared to those of airfoils, experimental data for finite wings lack variety and availability due to the added complexity in experiments. For example, a significantly higher number of pressure sensors is required to provide a thorough coverage of spanwise stations. The model for predicting three-dimensional pressure and lift coefficient must thus be able to develop a strong generalization capability from a sparser training data set.

\begin{table} 
\resizebox{\textwidth}{!}{
\begin{tabular}{l c c c c c c c}  
\toprule  
& \multicolumn{5}{c}{Number of unique entries per category} \\  
\cmidrule(l){3-6}
Model & Data Type & Sources & Airfoils/wings & Operating conditions & Individual measurements & $\alpha$ range & $M_\infty$ range \\
\midrule 
LAM (Lee et al.,~\citeyear{lee2024_LAM}) & Airfoil & 38 & 69 & 2917 & 109,020 & $-30^\circ - 30^\circ$ & $0.0-1.0$ \\  
Current work & Wing    & 16 & 16 & 54   & 23,723  & $-7^\circ - 12^\circ$  & $0.0-0.8$ \\  
\bottomrule  
\end{tabular}
}
\caption{Comparison between the wing database built for this study and the experimental airfoil $C_p$ database created previously for the Large Airfoil Model.} 
\label{tab:database_comparison}
\end{table}

\begin{figure}[h!]
     \centering
     \subfigure[Taper ratio, $\lambda$]{\includegraphics[width=0.95\textwidth]{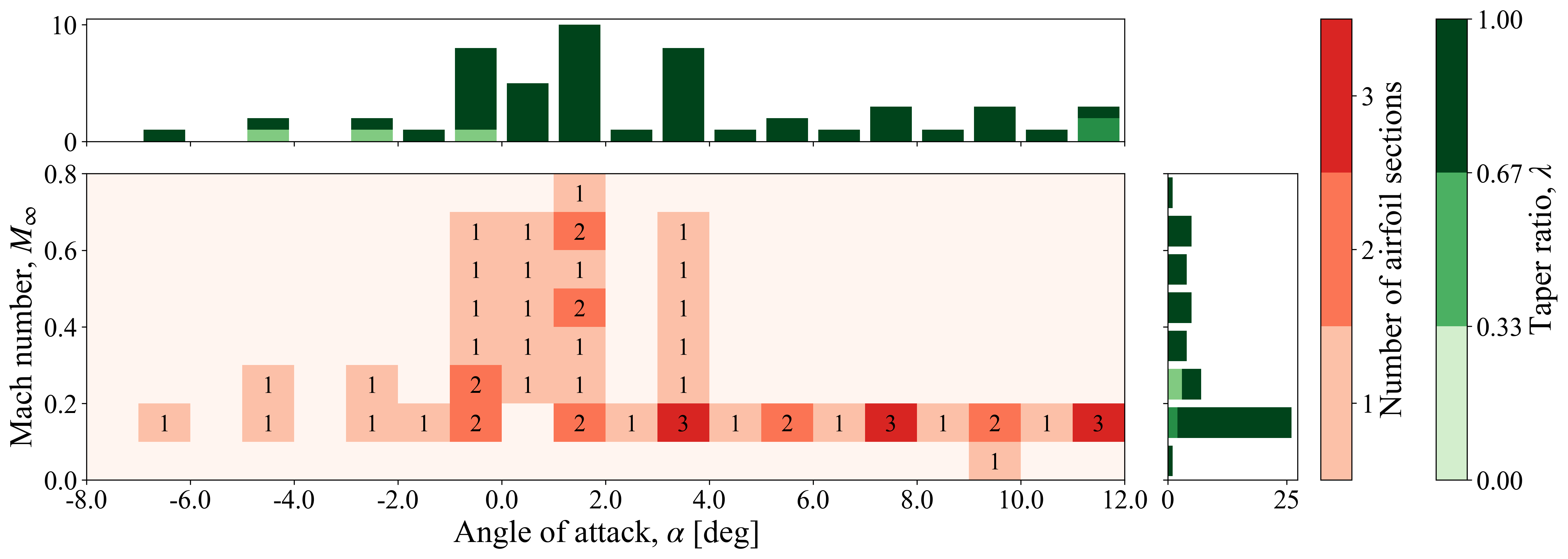}\label{fig:datadist_taper}}
     \subfigure[Leading edge sweep angle, $\Lambda$]{\includegraphics[width=0.95\textwidth]{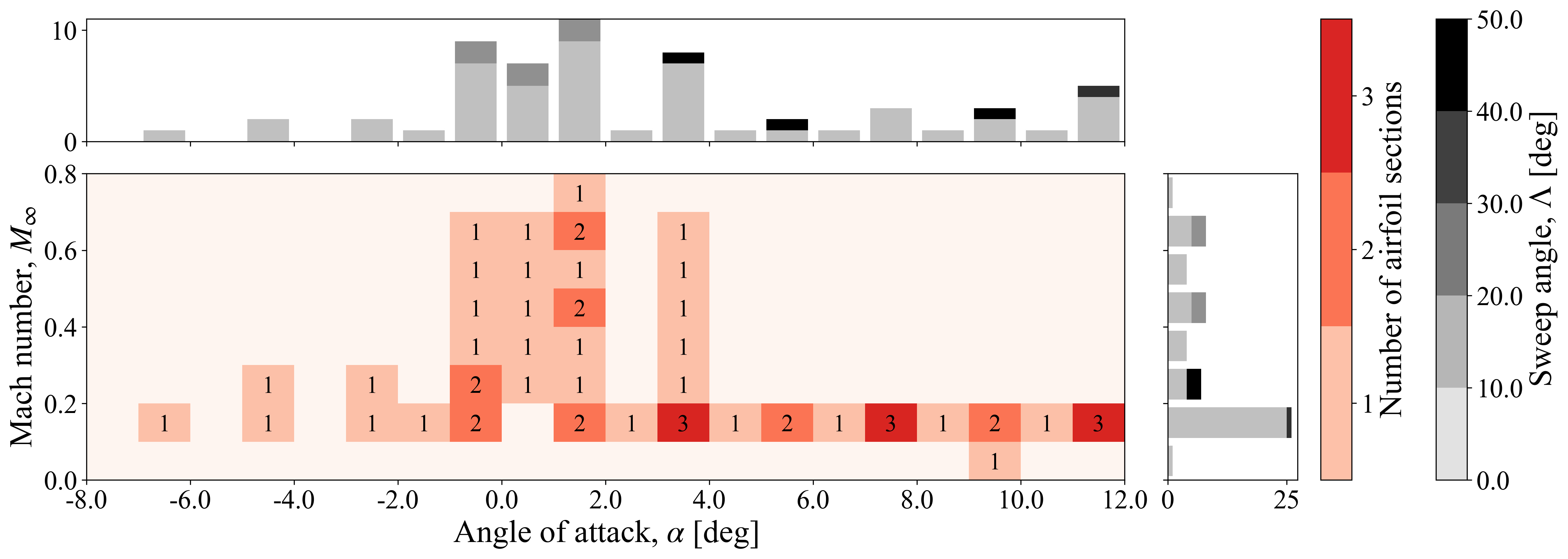}\label{fig:datadist_sweep}}
     \subfigure[Semi-span length ratio, $s/c_\text{root}$]{\includegraphics[width=0.95\textwidth]{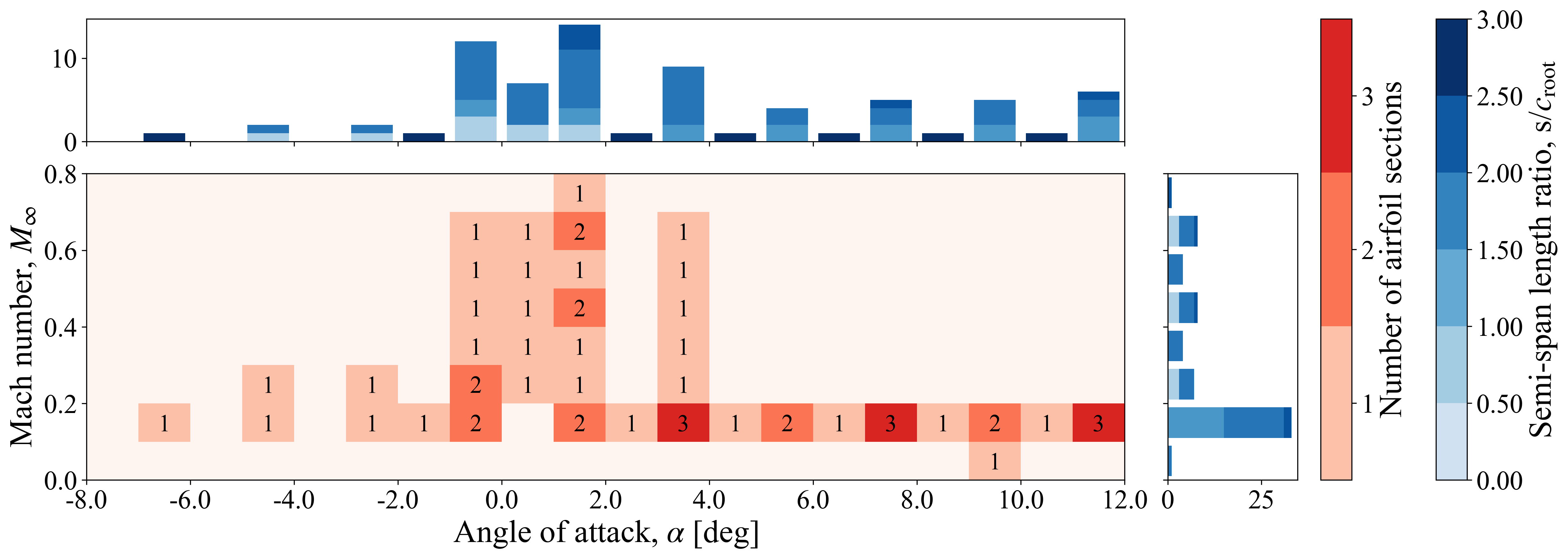}\label{fig:datadist_ar}}
     \caption{The distribution of available $C_p$ data with respect to the number of unique airfoil sections at each freestream Mach number and angle of attack. The marginal histograms are organized by different parameters that define the wing geometries}
     \label{fig:datadistribution}
\end{figure}

This comparative sparsity is presented in Fig.~\ref{fig:datadistribution} which visualizes how the database is distributed in the $\alpha$--$M_{\infty}$ space in terms of unique airfoil sections, taper ratio ($\lambda$), leading edge sweep angle ($\Lambda$), and semi-span/$c_\text{root}$ ($s/c_\text{root}$). Currently, the wing database includes a total of seven unique airfoil sections: NACA 0012, NACA 0015, NACA 0023, RAE 101, NACA 64A-105, a 5\% thick supercritical airfoil~(Bennett and Walker,~\citeyear{nasa1999209130}), and ``NACA 6-series-like airfoil''~(Soltani et al., \citeyear{soltani}). These selections are representative airfoils from both legacy and transonic applications.

\clearpage
The data distribution in terms of $M_\infty$ and $\alpha$ indicates that most digitized data points, based on available open-source experimental reports, are concentrated in the low subsonic regime ($M_\infty < 0.3$, or incompressible regime), particularly at small positive angles of attack ranging from approximately $0^\circ$ to $4^\circ$. The marginal histograms make this bias more evident; the majority of airfoil sections are concentrated around $M_\infty = 0.1$ to $0.3$ and $\alpha = 2^\circ$ to $4^\circ$. There is some coverage at higher Mach numbers, stemming from the inclusion of transonic in the database. This distribution likely reflects the fact that many experimental sources within the database are from 1970s to 1990s (such as the works by Spivey,~\citeyear{spivey1970}; Lockman and Seegmiller,~\citeyear{nasa84367}; Bragg,~\citeyear{nasa195330}). It was found that there was a heavier emphasis on conditions relevant to low-speed aerodynamic testing, during these times. This may be due to the fact that flow remains largely attached to the airfoil and remains in the incompressible regime, simplifying both experiments and analysis.

In terms of taper ratio, the marginal histogram in Fig.~\ref{fig:datadist_taper} reveals a strong bias in the current database toward a taper ratio of 1.0, or wings with equal $c_\text{root}$ and $c_\text{tip}$ (as defined in Fig.~\ref{fig:wing_geom_definition}). This bias stems from two main factors. First, rectangular wings are typically the simplest to manufacture due to their straightforward geometry. Consequently, when experimental testing of the three-dimensional characteristics of a specific airfoil is required, rectangular wings are often the default choice. Second, the digitization effort intentionally prioritized rectangular wings to ensure a sufficient volume of training data for the machine learning model developed in this work.

Figure~\ref{fig:datadist_sweep} also shows that the leading-edge sweep angles of wings in the database are primarily concentrated in the 0° to 10° range. As discussed in relation to the taper ratio, this trend arises from the database’s emphasis on rectangular or lightly tapered wings. Additionally, the distribution of sweep angles is closely correlated with the distribution of freestream Mach numbers, $M_\infty$. Higher sweep angles are typically employed at higher Mach numbers to reduce the effective Mach number incident on the wing surface~(Anderson, \citeyear{anderson}). However, as a substantial portion of the data corresponds to the incompressible flow regime, high sweep angles are generally unnecessary and are underrepresented.

Lastly, Fig.~\ref{fig:datadist_ar} demonstrates that the range of $s/c_\text{root}$ within the database is quite varied, with most falling in between 3.0 and 6.0. This distribution is consistent with the fact that the aspect ratios of conventional subsonic aircraft range typically from 6.0 to 8.0, a result of compromise between induced drag\footnote{Induced drag is a type of drag that occurs on a finite wing due to downwash, which reduces the effective angle of attack and tilts the lift vector backward, creating a drag component.} reduction and structural integrity~(Anderson, \citeyear{anderson}). This is equivalent to $s/c_\text{root}$ of 3.0 to 4.0 for rectangular wings. It is thus natural that this range is the most investigated in the wind tunnel, as observed in the database.

Our ongoing data mining and database expansion efforts aim to populate under-represented regions of the parameter space, thereby enabling more comprehensive training and validation of future machine-learning models. Improved coverage can be achieved through several approaches. One is targeted data mining, where future digitization efforts focus on identifying and extracting experimental data from studies that explore currently underrepresented areas. Additionally, since the database is publicly available, expansion can be facilitated through collaborations with experimental facilities and research groups to obtain unpublished or legacy datasets---especially those involving rare geometries or test conditions.

\subsection{Data Pre-processing}
Typically in literature, a $C_p$ distribution at a given spanwise station of a wing is reported as a function of the normalized coordinates in the chordwise direction, $x/c$. The one-dimensional parametrization can prove limiting, as $C_p$ is not one-to-one in $x/c$. A model trained on this input format would not be able to distinguish the suction and pressure sides of a wing. Another important property of $C_p$ curves is that the trailing edge pressures of each surface match, unless the wing is undergoing severe trailing edge separation. Thus, a different coordinate system must be used to enforce these conditions.
 
\begin{figure}[h!]
\centering
\includegraphics[width=.85\textwidth]{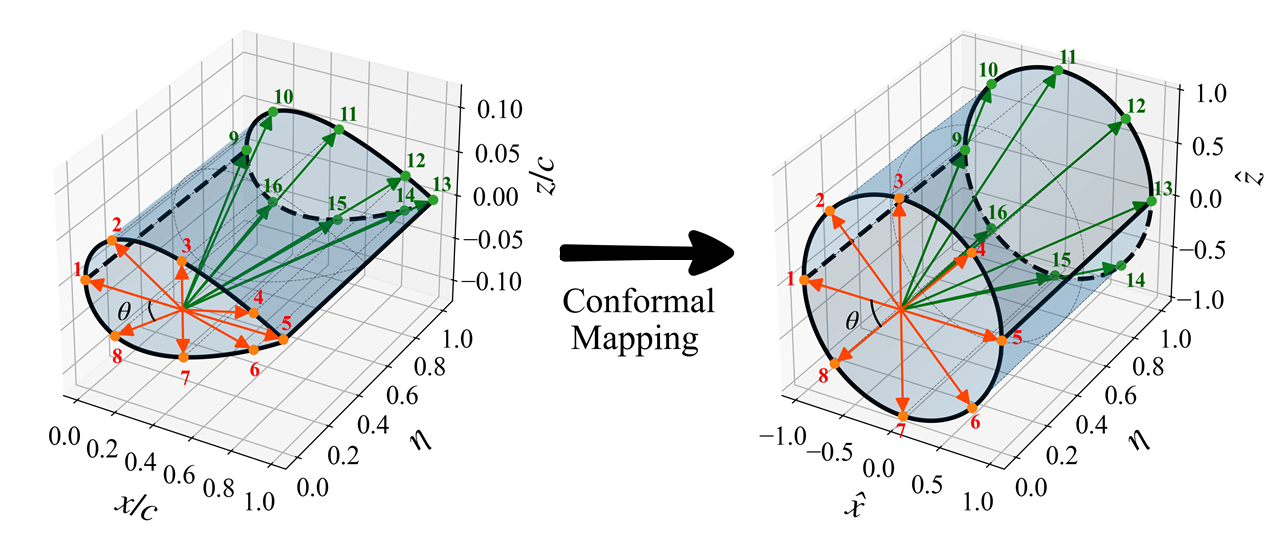}
\caption{Transformation of three-dimensional wing coordinates ($x/c$, $z/c$, $\eta$) to equivalent cylindrical coordinates ($\hat{x}$, $\hat{z}$, $\eta$) via conformal mapping of the airfoil section}
\label{fig:conf_mapping}
\end{figure}

It was previously found that a conformal mapping-inspired approach provides an effective means to satisfy these requirements by transforming the normalized airfoil chordwise location into $\hat{x}$ and $\hat{y}$ in a polar coordinate system~(Lee et al.,~\citeyear{lee2024_LAM}). In this work, the same approach is extended to a three-dimensional domain. The normalized chordwise direction $x/c = \{x \in \mathbb{R} \mid 0 \le x \le 1 \}$ is transformed linearly such that $\hat{x} = \{x \in \mathbb{R} \mid -1 \le x \le 1 \}$. The equivalent angle is obtained via $\theta = \cos^{-1}(x/c)$, meaning that $\hat{z} = \sin{\theta}$, and $\hat{z} = \{z \in \mathbb{R} \mid -1 \le z \le 1 \}$. The normalized spanwise location $\eta =  \{y \in \mathbb{R} \mid 0 \le y \le 1 \}$ is left unchanged from the original coordinate system. In this transformation, $\hat{x}$ and $\hat{z}$ can be interpreted as the $x$- and $z$-values of the unit cylinder, as visualized in Fig.~\ref{fig:conf_mapping}. The approach differentiates the upper and lower surfaces by the sign of $\hat{z}$, and enforces the periodicity at the leading and trailing edges by having the same $\hat{x}$ and $\hat{z}$ values at these locations.

From the database, a subset of the data consisting of exclusively rectangular wings was down-selected as the training data. The training data was then pre-processed to extract the input variables required for the model. These input variables included 28 $z$-coordinates of the section geometry ($\mathbf{z}/c$) at specified chordwise locations ($\mathbf{x}/c$) for the upper and lower surfaces, along with $\alpha$, $M_\infty$, $s/c_\text{root}$, $\lambda$, $\Lambda$, $\hat{x}$, $\hat{z}$, and the non-dimensional semi-spanwise location $\eta$. Together, this resulted in a total of 64 input variables. This approach is particularly well suited for data sets compiled from diverse experiments, as it explicitly defines the pressure measurement locations, which often vary across sources.

\section{Methodology: Bayesian Approach to Pressure and Lift Prediction}\label{sec:model}
\subsection{Gaussian Process Regression}
In the Gaussian process (GP) formalism adopted here, the data presented in Section~\ref{sec:data} are assumed to contain $N$ pairs of measurement locations and sensor readings $\left( \textbf{x}_i, f(\textbf{x}_i) \right)$. Predictions are sought at $M$ new locations, $\mathbf{x}_i^{\ast}$. For convenience, the inputs and outputs are collected into the matrices and vectors
\begin{equation}
    \textbf{X} = \begin{bmatrix} \textbf{x}_1\\ \vdots \\ \textbf{x}_N \end{bmatrix}, \;
    \textbf{y} = \begin{bmatrix} f(\textbf{x}_1)\\ \vdots \\f(\textbf{x}_N) \end{bmatrix}, \; \; \text{and} \; \; 
    \textbf{X}^\ast = \begin{bmatrix} \textbf{x}_1^\ast\\ \vdots \\ \textbf{x}_M^\ast \end{bmatrix}, \;
    \textbf{f}^\ast = \begin{bmatrix} f(\textbf{x}_1)\\ \vdots \\ f(\textbf{x}_M) \end{bmatrix}.
\end{equation}
As the training measurements are experimental, each observed value is assumed to be corrupted by an additive Gaussian noise $f \left(\mathbf{x}_{i} \right) = r(\mathbf{x}_i) + \epsilon_i, \; \epsilon \sim \mathcal{N} \left(\mathbf{0}, \boldsymbol{\Sigma_s} \right)$ where $r$ denotes the true (noise-free) pressure field and $\boldsymbol{\Sigma_s}$ is the noise covariance. This constitutes the likelihood model of the GP. With a zero-mean GP prior, $r \left( \cdot \right) \sim \mathcal{N}(\mathbf{0}, \boldsymbol{K})$; $\boldsymbol{K} = k(\mathbf{X}, \mathbf{X}^\prime)$, the posterior predictive distribution at test inputs $\mathbf{X}^{\ast}$, is given by
\begin{align}
\bm{\mu}_{*|\mathbf{X}} &= \mathbf{K}_{\mathbf{X}\ast}^T (\mathbf{K}_{\mathbf{X}\mathbf{X}} + \mathbf{\Sigma_s})^{-1}\mathbf{y}  \label{eq:posteriorMean}, \\
\mathbf{\Sigma}_{*|\mathbf{X}} &= \mathbf{K}_{**} - \mathbf{K}_{\mathbf{X}*}^T (\mathbf{K}_{\mathbf{X}\mathbf{X}} + \mathbf{\Sigma_s})^{-1} \mathbf{K}_{\mathbf{X}*}, \label{eq:posteriorCov}
\end{align}
\noindent where $\mathbf{K}_{\mathbf{X}\mathbf{X}} = k(\mathbf{X}, \mathbf{X})$, $\mathbf{K}_{\mathbf{X}\ast} = k(\mathbf{X}, \mathbf{X}^{\ast})$, and $\mathbf{K}_{\ast\ast} = k(\mathbf{X}^\ast, \mathbf{X}^\ast)$.

A direct inversion of $(\mathbf{K}_{\mathbf{X}\mathbf{X}} + \mathbf{\Sigma_s})^{-1}$ can be numerically unstable, particularly for large $N$. As is standard practice, a Cholesky factorization ($\mathbf{K}_{\mathbf{X}\mathbf{X}} + \mathbf{\Sigma_s} = \mathbf{L}\mathbf{L}^T$) is employed, where $\mathbf{L}$ is triangular. Substituting this factorization into Eqs.~\ref{eq:posteriorMean} and \ref{eq:posteriorCov} and solving the requisite triangular systems yields the numerically robust forms
\begin{align}
\bm{\mu}_{*|\mathbf{X}} = \mathbf{K}_{\mathbf{X}*}^T \boldsymbol{\alpha}\label{eq:posteriorMeanChol}, \text{ and} \\
\bm{\Sigma}_{*|\mathbf{X}} = \mathbf{K}_{**} - \mathbf{v}^T\mathbf{v},\label{eq:posteriorCovChol}
\end{align}
\noindent where $\boldsymbol{\alpha} = \mathbf{L}^{-T} \left( \mathbf{L}^{-1} \mathbf{y} \right)$ and $\mathbf{v} = \mathbf{L}^{-1}  \mathbf{K_{\mathbf{X}*}}$. These expressions avoid explicit matrix inversion, improve numerical stability, and leverage triangular solves readily available in standard numerical linear algebra libraries.

\subsection{Deep Kernel Learning Model}
The model proposed in this study is a modified deep kernel learning (DKL) model, with its architecture inspired by Wilson et al.,~\citeyearpar{Wilson_2016}, who first introduced the modeling framework, and by Lee et al.,~\citeyearpar{lee2024_LAM}, who successfully employed the architecture to predict $C_p$ distributions for two-dimensional airfoils. It was found that a DKL model is well suited to an unstructured data set, a natural consequence of aggregating various experimental data from different sources.

In this model architecture, a neural network is first used to create a complex, non-linear mapping from the original input space to a new latent space. A Gaussian process model is built in this latent space to make predictions from the extracted features. Being fundamentally a Bayesian approach, the resulting prediction is a multivariate Gaussian distribution over the space of $C_p$ values. 

\begin{figure}[h!]
\centering
\includegraphics[width=.75\textwidth,clip]{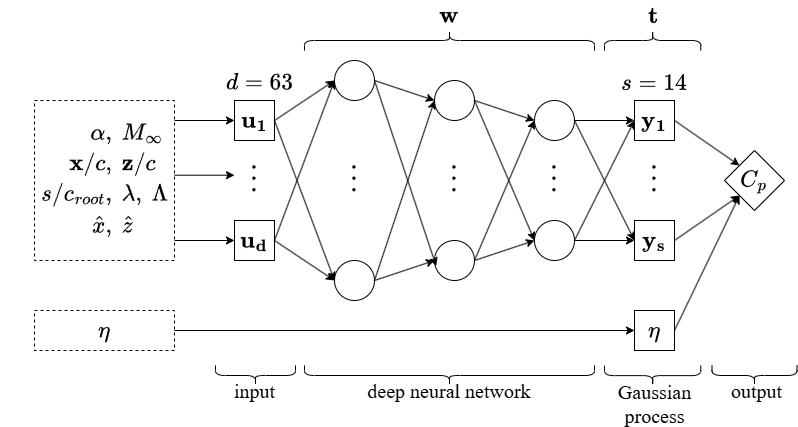}
\caption{Overview of the modified deep kernel learning model that maps a 64-dimensional training input to a 15-dimensional active space from which a Gaussian process model is built}
\label{fig:DKL_diagram}
\end{figure}

However, it was found that utilizing a purely latent space resulted in a loss of information regarding wing spanwise geometry, where the input variable of $\eta = 1.0$ did not necessarily equate to the wing tip in the latent space. The model performance was negatively impacted by this, especially in capturing the influence of the tip vortex. To address this, $\eta$ was directly input into the Gaussian process layer, enabling the model to explicitly enforce the constraint $0.0 \le \eta \le 1.0$; the process can be defined as
\begin{equation}
\begin{split}
    C_p \left(\underbrace{\alpha, M_\infty,}_{\textrm{\shortstack{operating \\ conditions}}} \underbrace{\mathbf{x}/c, \mathbf{z}/c,}_{\textrm{\shortstack{airfoil \\ geometry}}}
    \underbrace{s/c_\text{root}, \lambda, \Lambda}_{\textrm{\shortstack{wing \\ geometry}}}
    \underbrace{\hat{x}, \hat{z}, \eta}_{\textrm{coordinates}} \right) & \equiv C_p \left( \underbrace{\mathbf{u}}_{\textrm{\shortstack{all neural \\ network inputs}}}, \eta\right) \sim \mathcal{N} \left( \mu \left( \mathbf{y}, \eta, \mathbf{t} \right), \Sigma\left( \mathbf{y}, \eta, \mathbf{t} \right) \right),  \\
\textrm{where} \; \; \underbrace{\mathbf{y}}_{\textrm{latent variable}} & = \underbrace{f_{\mathbf{w}} \left(\mathbf{u} \right)}_{\textrm{deep neural network with weights} \; \mathbf{w}},
\end{split}
\end{equation}

\noindent where the notation $f_{\mathbf{w}}: \mathbf{u} \rightarrow \mathbf{y}$ denotes the deep neural network that is parameterized with weights $\mathbf{w}$. The space $\mathbf{u} \in \mathbb{R}^{d}$ comprises the operating conditions, wing section (airfoil) geometry, wing planform geometry, and \emph{conformal} airfoil coordinates, i.e. $\mathbf{u} = \left\{ \alpha, M_\infty, \mathbf{x}/c, \mathbf{z}/c, s/c_{\text{root}}, \lambda, \Lambda, \hat{x}, \hat{z}\right\}$. The outputs of the deep neural network are latent variables $\mathbf{y} \in \mathbb{R}^{s}$. In the latent space defined by $\mathbf{y}$ and $\eta$, a Gaussian process model is built. The model is defined by a mean and covariance function, which is based on a two-point kernel function $k_{\mathbf{t}}$, where the subscript $\mathbf{t}$ indicates hyperparameters to parameterize the kernel function. The overall modified DKL architecture of the model is presented in Fig.~\ref{fig:DKL_diagram}. 

The deep neural network component of the model has a [63-1000-1000-500-50-14] architecture that maps the space of inputs $\mathbf{u}$ to the latent space defined by $\mathbf{y}$. It was previously found that 60 input variables and 10 output dimensions were sufficient to capture the variations in $C_p$ of an airfoil~(Lee et al.,~\citeyear{lee2024_LAM}). The addition of 4 dimensions, including the non-dimensional spanwise variable $\eta$, was found to be sufficient to model the three-dimensional effects in a finite wing. 

For the GP layer, a multi-dimensional product kernel was used. Defined as the Hadamard (element-wise) product between kernels for each latent variable, the equation for the kernel is 

\begin{equation}
 k(\mathbf{X},\ \mathbf{X^\prime}) = k_1(\mathbf{y_1},\ \mathbf{y_1^\prime}) \odot k_2(\mathbf{y_2},\ \mathbf{y_2^\prime}) \odot \hdots  \odot k_s(\mathbf{y_s},\ \mathbf{y_s^\prime}) \odot k_{\eta}(\eta,\ \eta^\prime),
\end{equation}

\noindent where the symbol $\odot$ represents an element-wise product between matrices. 

For the kernel function in each $\mathbf{y}$ dimension ($k_1, \ \dots,\  k_s$), the Mat\a'ern 5/2 function was used. For $k_\eta$, the Mat\a'ern 3/2 function with spatially varying hyperparameters (further detail in Section~\ref{subsec:svh}) resulted in the best model accuracy. The Mat\a'ern family of kernels allows the model to more accurately capture aerodynamic phenomena on lifting surfaces, such as the pressure changes at the suction peak and three-dimensional effects induced by tip vortices. As these phenomena are typically localized and have sharp gradients, the overly smooth nature of commonly used kernels such as the squared exponential makes them less suitable for capturing such effects accurately. The equations for the Mat\a'ern functions are 

\begin{align}
k_i(\mathbf{y}_i, \mathbf{y}_i^\prime) &= \sigma_{f,i}^2 \left(1 + \frac{\sqrt{5}}{\ell_i}\dist{\mathbf{y}_i - \mathbf{y}^\prime_i} + \frac{5}{3\ell_i^2}\dist{\mathbf{y}_i - \mathbf{y}_i^\prime}^2\right) \exp\left(-\frac{\sqrt{5}}{\ell_i}\dist{\mathbf{y}_i - \mathbf{y}_i^\prime}\right),\\ 
k_\eta(\eta, \eta^\prime) &= \sigma_{f,\eta}^2\left(\eta\right) \left(1 + \frac{\sqrt{3}}{\ell_\eta(\eta)}\dist{\eta - \eta^\prime}\right) \exp\left(-\frac{\sqrt{3}}{\ell_\eta(\eta)}\dist{\eta - \eta^\prime}^2\right),\label{eq:matern}
\end{align}

\noindent where $i=1, \ldots, s$, and $\sigma_{f,\cdot}^2, \ \ell_\cdot$ are the model hyperparameters: the signal variance and the characteristic length scale. The subscripts indicate the corresponding kernel for each hyperparameter. 

A GP model requires the definition of the likelihood model, typically through a Gaussian noise model that accounts for measurement or modeling uncertainty. For aerodynamic experimental data, the noise model can typically be set to the sensor noise as presented in the source document (i.e. $\sigma^2_{\text{expt}}\mathbf{I}$), provided that the original authors quantified the measurement noise. However, it was not uncommon to encounter wind tunnel measurement data lacking uncertainty quantification---particularly in older experiments, where limitations in the experimental apparatus were more pronounced. As a result, some measurements exhibited highly noisy behavior that cannot be systematically accounted for. Furthermore, additional noise is inevitably introduced when digitizing individual data points from graphical data. These facts made it difficult to define a fixed heteroscedastic noise model. A separate hyperparameter, $\sigma^2_{\text{add}}$, was introduced to represent any additional noise not explicitly quantified in the training data. The total likelihood model is thus defined as $\boldsymbol{\Sigma}_{\mathbf{s}} = \sigma^2_{\text{expt}}\mathbf{I} + \sigma^2_{\text{add}}\mathbf{I}$. 

Although experimental data was used in this model---primarily due to the clearly defined uncertainty values---the Deep Kernel Learning architecture is highly flexible. The model is not restricted to experimental data alone. A data set consisting solely of computational data, or even a multi-fidelity dataset combining experimental and computational sources, can be accommodated within the same architecture without modifications.

\subsection{Spatially Varying Hyperparameters}\label{subsec:svh}
\begin{figure}[h!]%
\centering
\subfigure[GP model with a smaller length scale, $\ell^2=0.018$]{%
\label{fig:small_ell}%
\includegraphics[width=.4\textwidth]{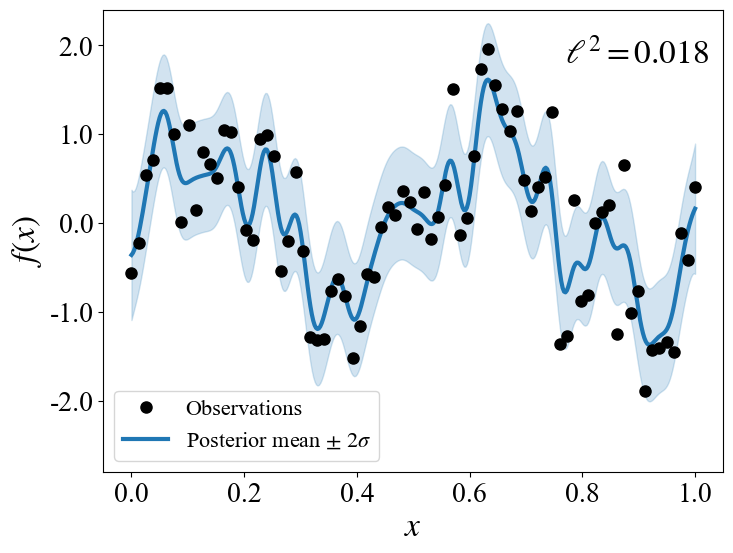}}%
\qquad
\subfigure[GP model with a larger length scale, $\ell^2=0.044$]{%
\label{fig:large_ell}%
\includegraphics[width=.4\textwidth]{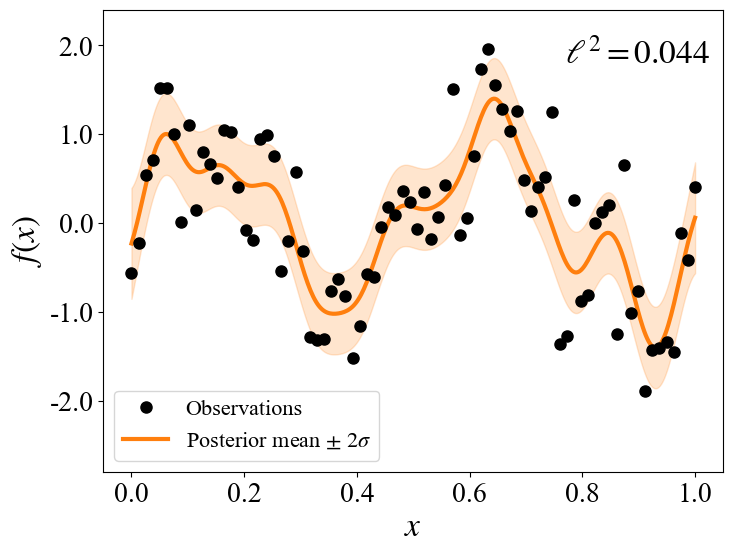}}%

\caption{Effect of length scale value on GP posterior predictive distribution on a synthetic data set}\label{fig:lengthscale_visualization}
\end{figure}

Since a Deep Kernel Learning model is fundamentally a Gaussian process (GP) model, its performance can be adjusted by tuning the GP hyperparameters. One such hyperparameter is the length scale, which can informally be thought of as the distance traveled in input space before the function value can change significantly~(Rasmussen and Williams,~\citeyear{RW2006}). Changes in the length scale have a significant impact on the smoothness of the Gaussian process model’s fit. Figure~\ref{fig:lengthscale_visualization} illustrates this property using a synthetic data set, where GP models are trained with varying values of the length scale. As shown, smaller length scales enable the model to capture high-frequency variations more closely, while larger length scales result in smoother predictions that emphasize broader trends at the expense of misrepresenting local fluctuations in the data.

For a pressure distribution over a wing, it is clear that the optimal length scale value depends on the spanwise location. Near the wing tip, the chordwise pressure distribution undergoes a localized distortion due to tip vortex effects~(McAlister and Takahashi,~\citeyear{nasa3151}). In the affected region, the length scale needs to be small to capture the changes accurately. Farther from the wing tip, the vortex effects are less pronounced and changes in the chordwise $C_p$ distributions are markedly more gradual. A data set of slowly changing observations benefits from greater length scales to prevent overfitting. 

It is thus necessary to capture the competing effects in a single model while minimizing posterior variance, which would lead to greater confidence in predictions and more reliable aerodynamic assessments of finite wings. To this effect, a kernel with spatially varying hyperparameters was adopted for the DKL model's kernel function in the spanwise ($\eta$) domain, as introduced in Eq.~\ref{eq:matern}. In this modified kernel function, the length scale and the signal variance are functions of the spanwise location along the wing (i.e. $\ell(\eta), \sigma_f^2(\eta))$. The wing was divided into two sections demarcated by the critical spanwise location, $\eta_\text{crit}$. The spanwise location below and above $\eta_\text{crit}$ can be thought of as the inboard and outboard regions, respectively; in the outboard region the effect of the tip vortex and other three dimensional phenomena dominate. For each of the sections, the length scale and the signal variance were inferred during the training process. To improve numerical stability when transitioning between regions, the shift from one value to another was smoothed using the logistic function. The function can be recast as

\begin{align}
f(\eta) = v_\text{inboard} + \frac{v_\text{outboard} - v_\text{inboard}}{1 + e^{-R(\eta - \eta_\text{crit})}} \label{eq:gen_log_fx},
\end{align}

\noindent where $v$ is the desired hyperparameter value and $R$ is the growth rate. It was found that $\eta_\text{crit} = 0.70$ and $R = 15$ yielded a favorable trade-off between prediction accuracy and numerical stability. 

\begin{figure}[h!]%
\centering
\subfigure[Length scale, $\ell$]{%
\label{fig:ell}%
\includegraphics[width=.4\textwidth]{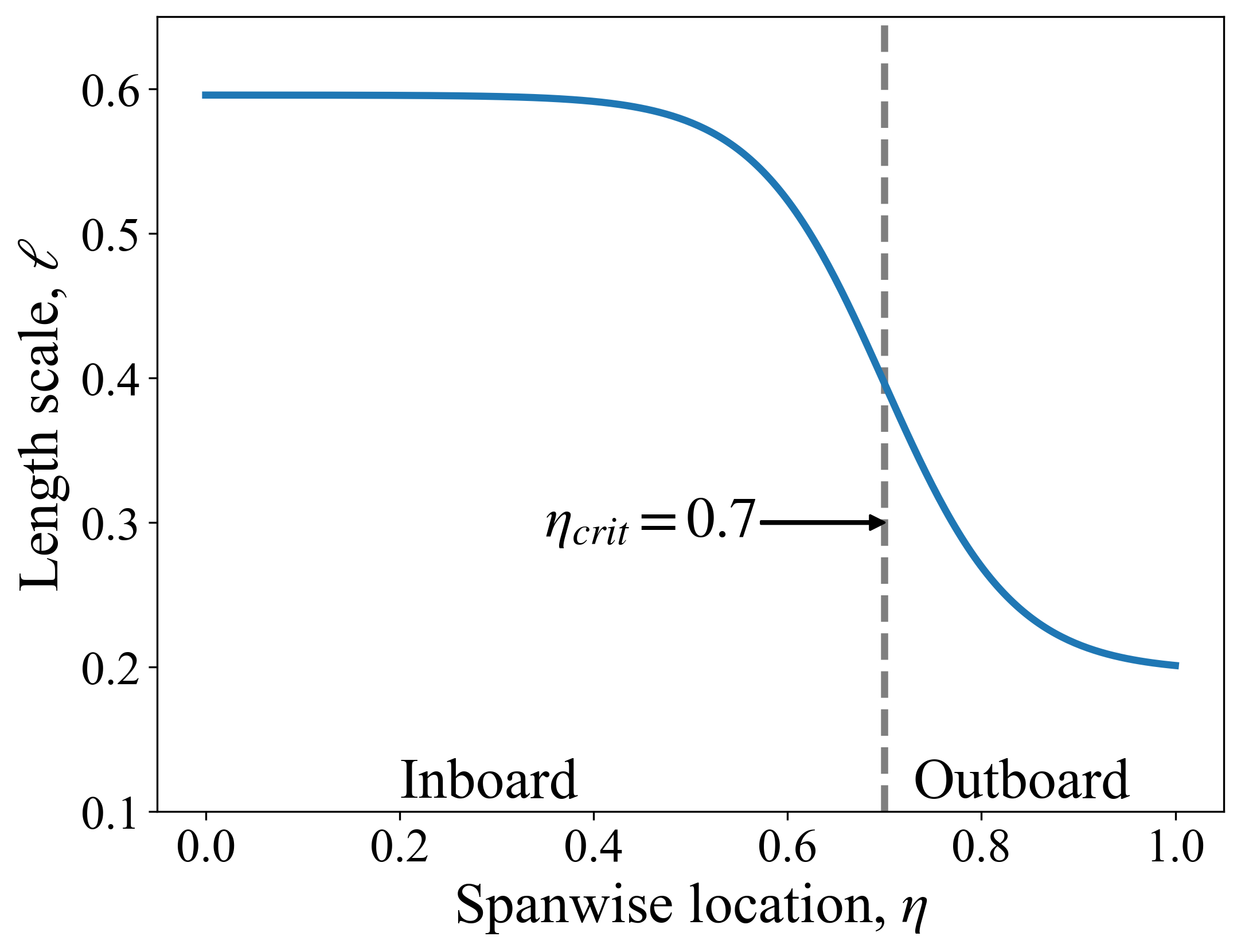}}%
\qquad
\subfigure[Signal variance, $\sigma_f^2$]{%
\label{fig:scale}%
\includegraphics[width=.41\textwidth]{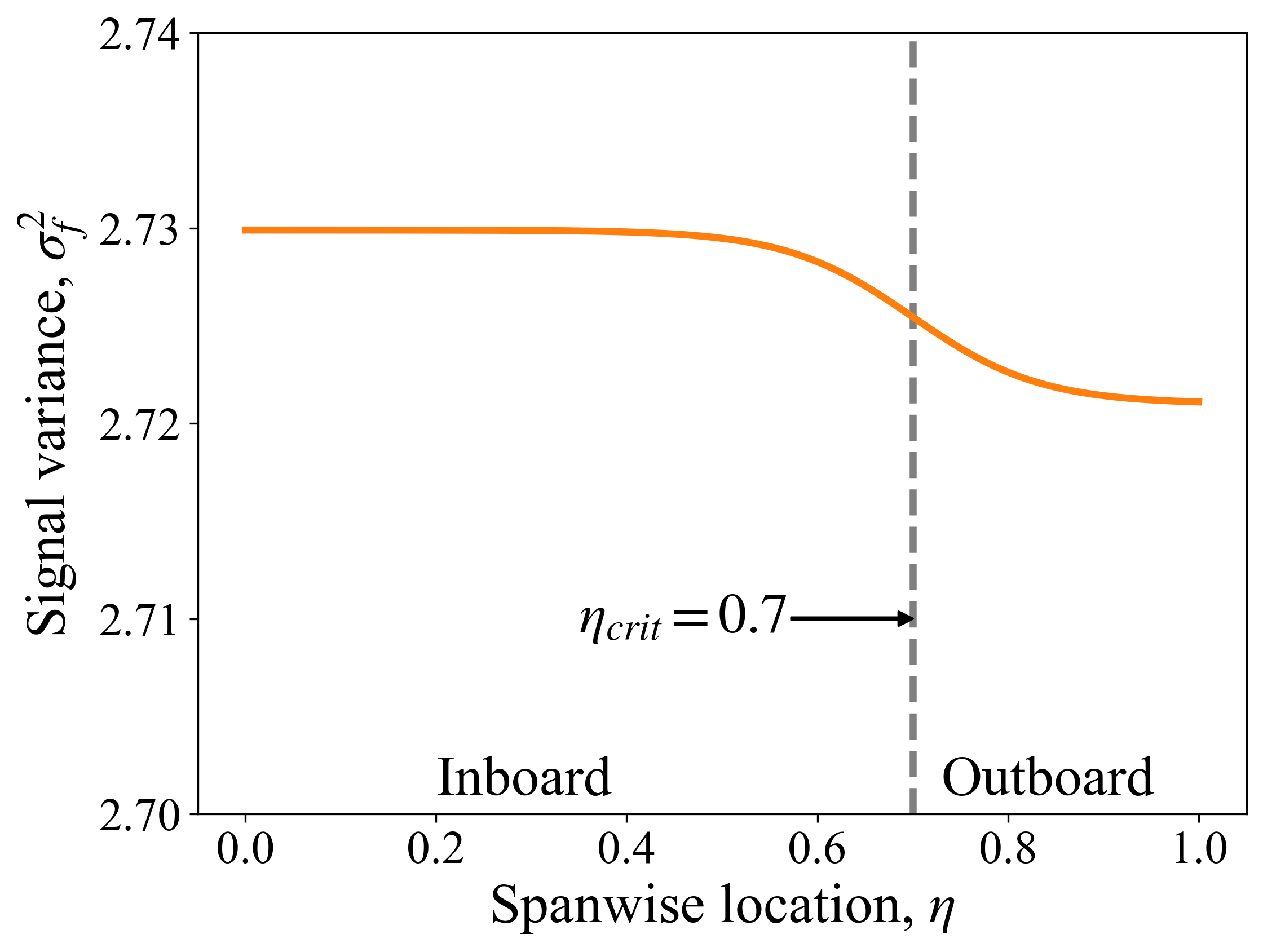}}%
\caption{Hyperparameter values with respect to the spanwise location for the spatially varying kernel}\label{fig:spatially_varying_hyperparam}
\end{figure}

Figure~\ref{fig:spatially_varying_hyperparam} demonstrates how the optimized length scale and signal variance vary along the spanwise direction in the final model. In the inboard section, the length scale (Fig.~\ref{fig:ell}) begins at 0.5958 but drops to 0.1967 in the outboard region due to the strong localized influence of the wing tip vortex on $C_p$. In contrast, the signal variance (Fig.~\ref{fig:scale}) exhibits only a minimal change, decreasing slightly from 2.7299 to 2.7210. This suggests that while the overall magnitude of the predicted variations remains consistent across the span, the model adjusts its sensitivity to local fluctuations by modulating the length scale.

\subsection{Large Airfoil Model Prior}
\begin{figure}[h!]%
\centering
\subfigure[GP prior distribution]{%
\label{fig:constMean_prior}%
\includegraphics[width=.4\textwidth]{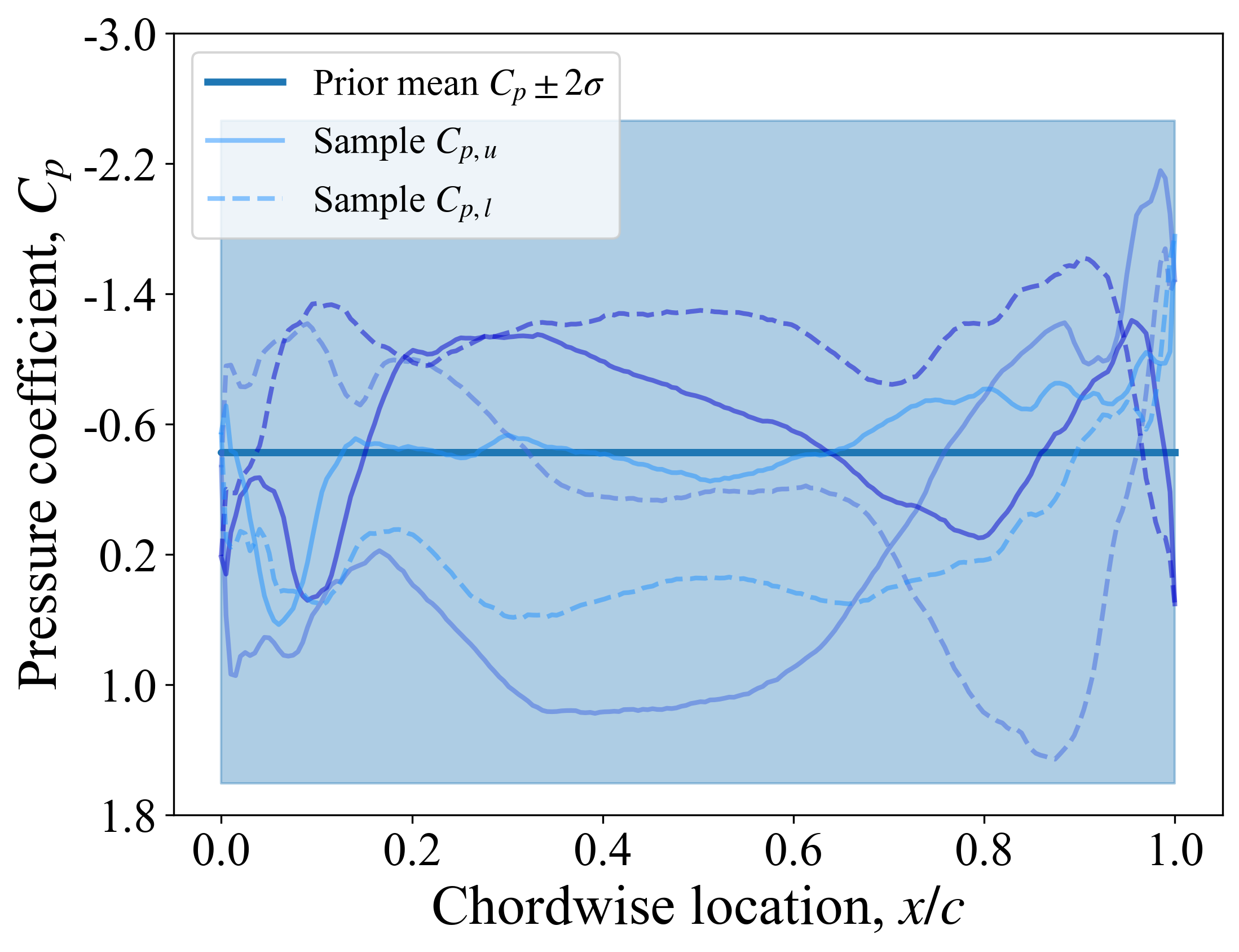}}%
\qquad
\subfigure[GP posterior predictive distribution]{%
\label{fig:constMean_posterior}%
\includegraphics[width=.4\textwidth]{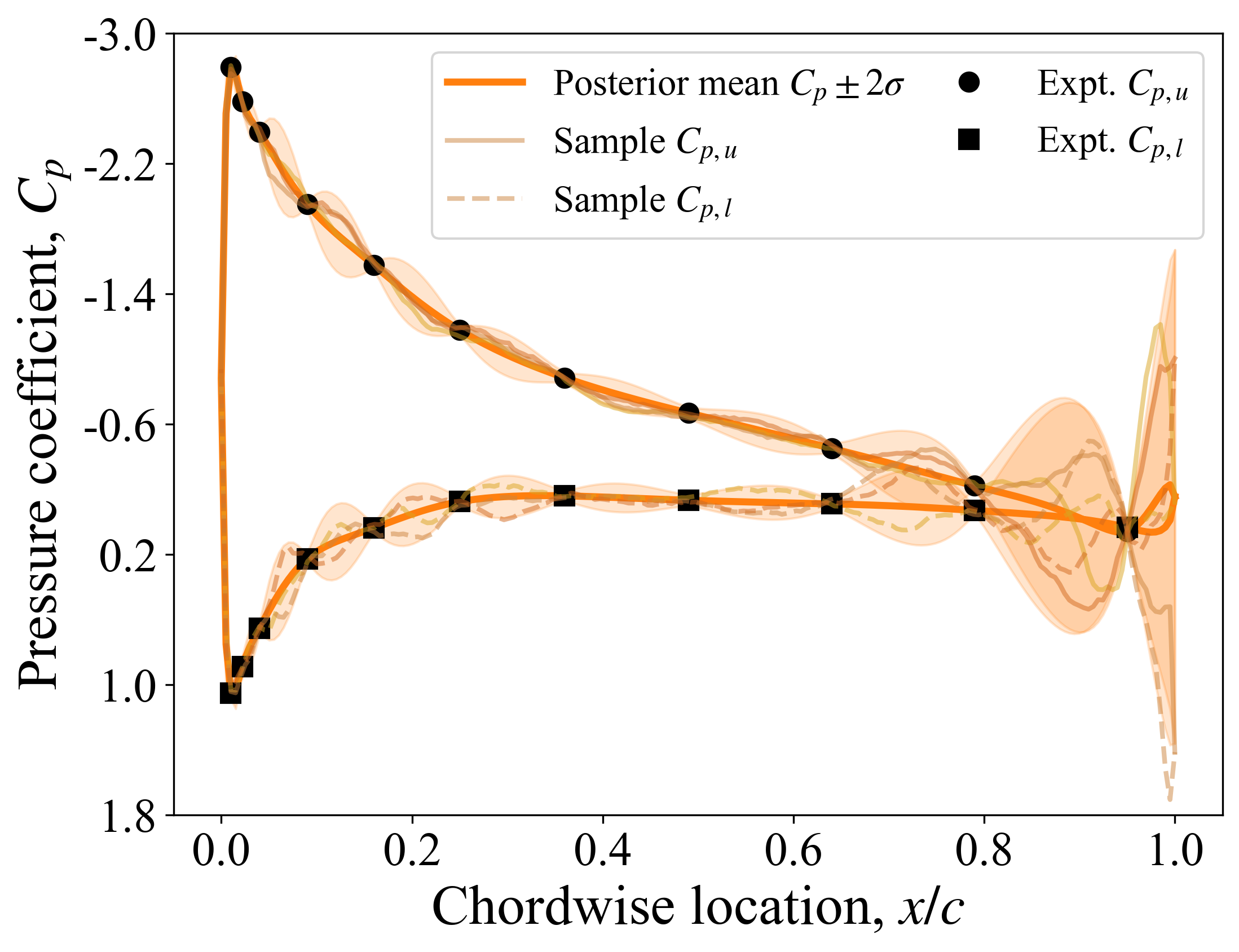}}%
\caption{Limitation of a constant mean prior in modeling of an airfoil $C_p$ distribution. The Mat\'ern 5/2 function was used as the kernel function for the GP model. The experimental training data were obtained from McAlister and Takahashi,~\citeyear{nasa3151}}\label{fig:constMean_limitation}
\end{figure}

When no observations inform the posterior, the Gaussian-process predictive mean simply reverts to the prior mean function, reflecting the model’s default belief in the absence of data. A typical prior mean function is a constant function (including the zero function), as visualized in Fig.~\ref{fig:constMean_prior}. However, the flatness of the prior makes it unsuitable when performing a GP regression on an aerodynamic measurements. This is especially true for pressure measurements, as it is common to have missing measurements at the test article trailing edges due to spatial limitations. When trained on these measurements, the predicted $C_p$ can coalesce toward the mean value at the trailing edge. The consequent upward inflection would be a non-physical airfoil behavior. Figure~\ref{fig:constMean_posterior} captures the posterior predictive distribution of a GP model with a constant mean prior trained on a real experimental $C_p$ distribution. The model's limitation is clearly demonstrated via non-physical $C_p$ ``loop'' at the trailing edge. Furthermore, the erroneous prediction is accompanied by a sizable confidence interval due to the large variance of the prior distribution necessary to capture the entire $C_p$ curve.

For experimental measurements of the wing $C_p$ distribution, the limitations are two-fold: measurements are taken at only a limited number of spanwise stations, and the chordwise resolution may not be comprehensive. As illustrated in Fig.~\ref{fig:constMean_limitation}, a GP model trained on such data can produce ``flat'' $C_p$ distributions at certain spanwise or chordwise locations. In other words, a suboptimal choice of the prior function can hinder the model's ability to accurately capture the underlying aerodynamic behavior.

Intuitively, this limitation can be circumvented by fitting the experimental data from a prior distribution that resembles the shape of an airfoil $C_p$ distribution rather than a flat line. Such incorporation of the airfoil physics into the model prior distribution is introduced in this paper, dubbed the Large Airfoil Model-prior (LAM-prior). The construction of the LAM-prior largely follows the workflow introduced by Seshadri et al.~\citeyearpar{seshadri2023}, and utilizes the Large Airfoil Model~(Lee et al.,~\citeyear{lee2024_LAM}) as the prior distribution generator.

In the model, the existence of a chordwise $C_p$ distribution at each spanwise station is assumed. This distribution is sourced from the Large Airfoil Model (LAM). The LAM is a machine learning model that generates uncertainty-aware predictions of airfoil $C_p$ distributions, given the airfoil geometry, $\alpha$, and $M_\infty$ as its inputs. It was trained on experimental data and incorporated the corresponding measurement uncertainties during inference. As a probabilistic model, its predictions are inherently multivariate normal distributions, which makes the model suitable as a prior distribution generator. 

The LAM's predictive posterior distribution is characterized by a mean and covariance. However, the LAM assumes that the pressure distribution is purely two-dimensional. In a flow over a wing, various three-dimensional effects, including the downwash induced by tip vortices and spanwise flow, make this assumption invalid. To generate a prior more suitable for capturing the increased flow complexity, a new LAM-prior distribution was created by averaging the predictive posterior distributions at different angles of attack between the wing geometric angle of attack ($\alpha_\text{geo}$) and $0^\circ$. The LAM-prior distribution is characterized by a multivariate normal distribution of $\mathcal{N} \left(\bm{\mu}^{\text{LAM}}, \boldsymbol{\Sigma}^\text{LAM}\right)$ where 

\begin{align}
\boldsymbol{\mu}^\text{LAM}(y, \eta \mid \mathbf{t}) &= \frac{1}{4} \sum_{n=1}^{4} \boldsymbol{\mu}^{\text{2D}}(\alpha_n), \label{eq:lam_prior_mean} \\
\boldsymbol{\Sigma}^\text{LAM}(y, \eta \mid \mathbf{t}) &= \frac{1}{4} \sum_{n=1}^{4} \boldsymbol{\Sigma}^{\text{2D}}(\alpha_n). \label{eq:lam_prior_cov}
\end{align}

\noindent Here, both $\bm{\mu}^{\text{LAM}}$ and $\boldsymbol{\Sigma}^\text{LAM}$ are functions of the chordwise and spanwise positions---indirectly through $y$ and directly through $\eta$, respectively---given the hyperparameters $\mathbf{t}$. $\bm{\mu}^{\text{2D}}(\alpha_n)$ and $\boldsymbol{\Sigma}^{\text{2D}}(\alpha_n)$ denote the mean and covariance of the raw two-dimensional $C_p$ predictions by the LAM. These predictions are generated at a series of angles of attack, $\alpha_n$, where $\alpha_n = \alpha_\text{geo}\frac{n-1}{3}$. 

This averaging process was inspired by Prandtl's lifting line theory, a classical model of finite wing aerodynamics. The theory, which prescribes an infinite number of horseshoe vortices distributed along a finite wing, results in a downwash distribution that alters the \textit{effective} angle of attack at each spanwise station. Since the effective angle of attack varies from the nominal wing angle of attack to zero over the entire span, the averaging operation in Eqs.~\ref{eq:lam_prior_mean} and \ref{eq:lam_prior_cov} can be interpreted as the mean $C_p$ distribution over the wing. This would provide the prior with sufficient flexibility to capture the variation in the effective angle of attack due to three-dimensional effects.

It was indicated that the LAM-prior is computed by averaging over four discrete angles of attack. As the number of sampled angles increases (approaching infinity), the result converges toward the ``true'' mean $C_p$. To determine a sufficient number of sample angles, a convergence study was conducted, using the NACA 0012 airfoil at a geometric angle of attack of 8 degrees, which provided a sufficient variety in $C_p$ at the intermediate angles. In this study, it was found that utilizing four angles yields an percentage error in the enclosed area of approximately 1\% compared to the close-to-infinite case of $n=20$, which was deemed sufficiently accurate for the purpose of constructing the prior.

\begin{figure}[h!]%
\centering
\subfigure[LAM-prior distribution]{%
\label{fig:lam_prior}%
\includegraphics[width=.4\textwidth]{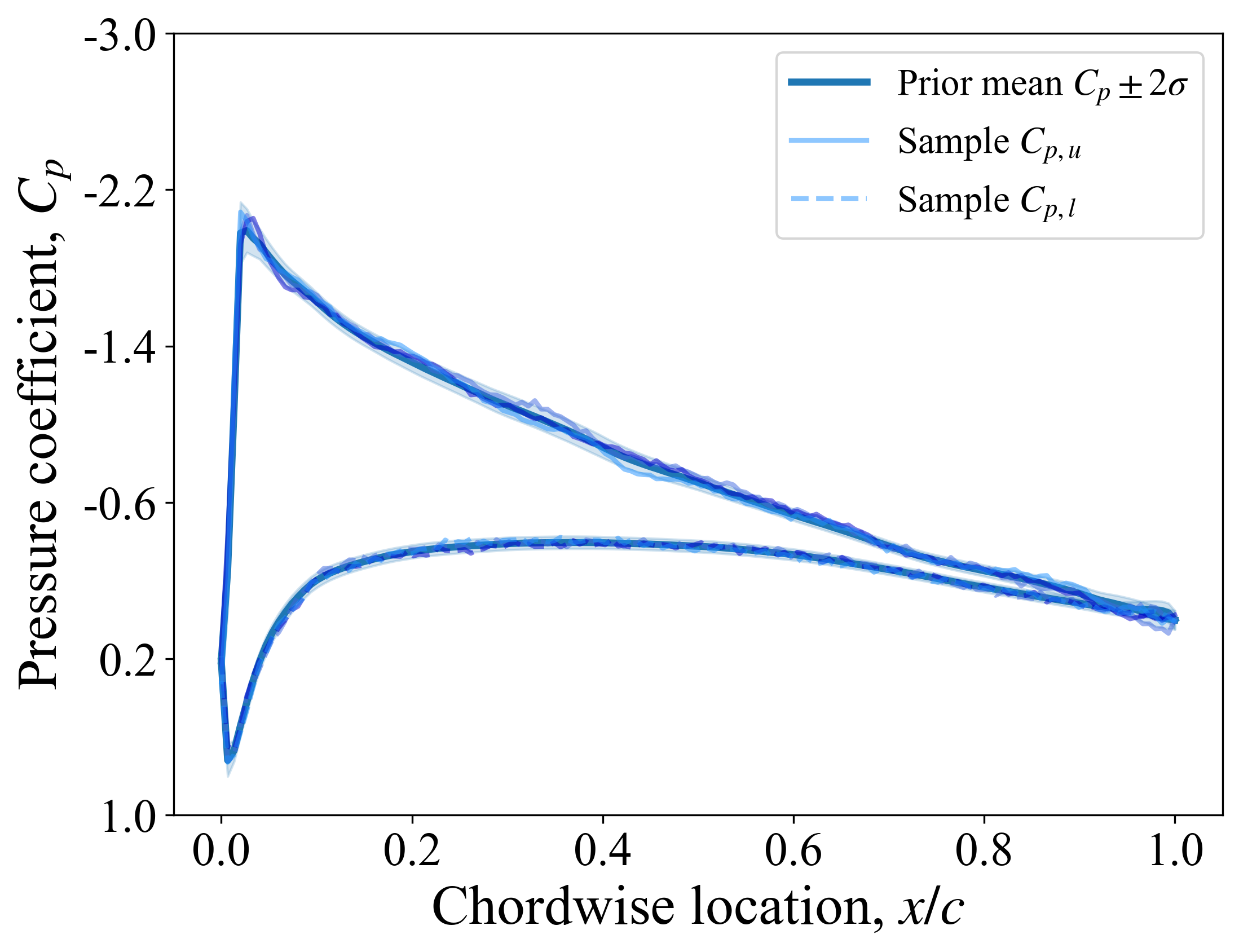}}%
\qquad
\subfigure[LAM-prior-informed posterior predictive distribution]{%
\label{fig:lam_posterior}%
\includegraphics[width=.4\textwidth]{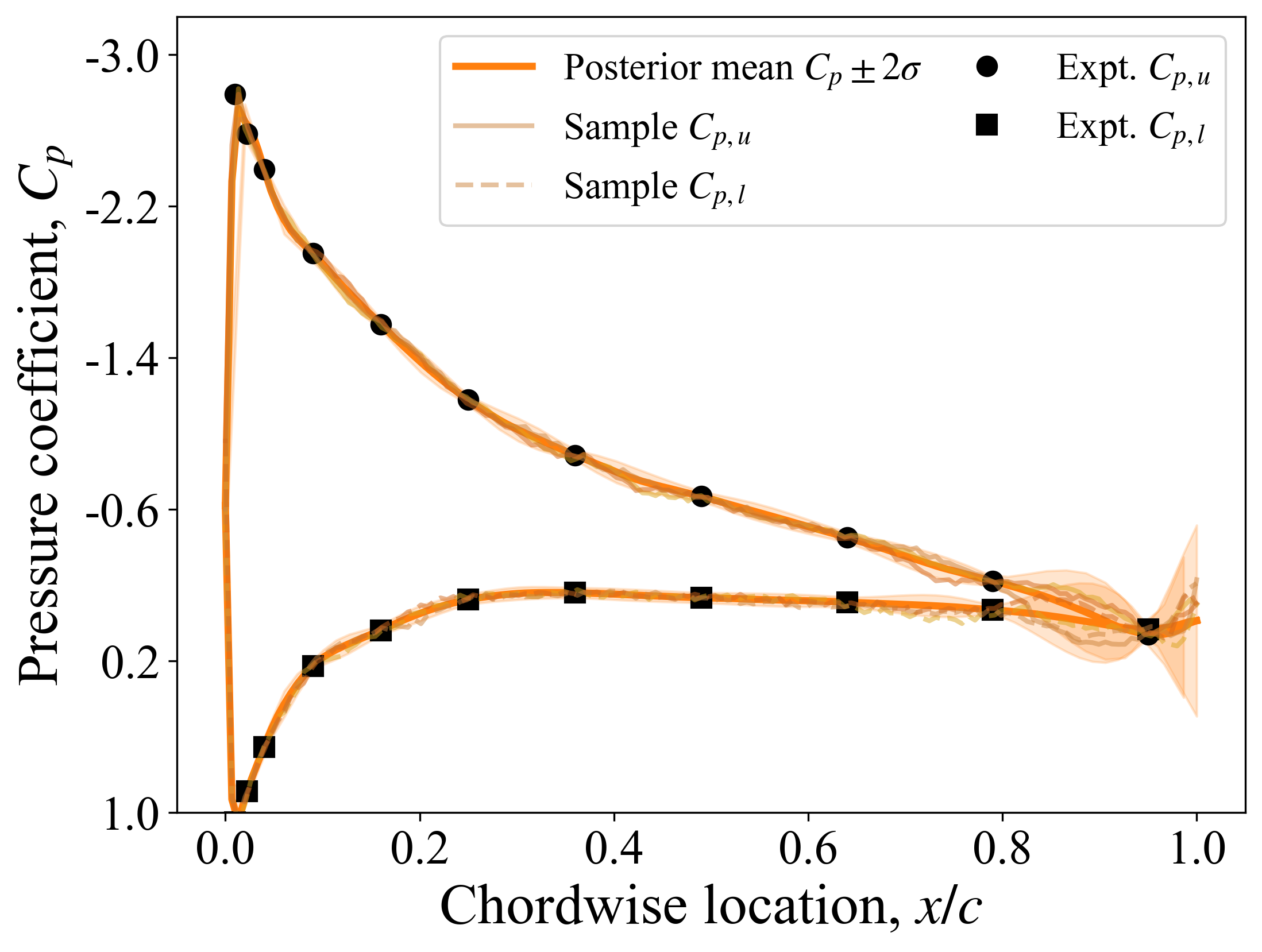}}%
\caption{The effect of utilizing a LAM-prior in modeling an airfoil $C_p$ distribution, demonstrating a reduction in posterior variance and the elimination of the trailing edge ``loop.'' The Mat\'ern 5/2 function was used as the kernel function for the GP model. The experimental training data were obtained from McAlister and Takahashi,~\citeyear{nasa3151}}\label{fig:lamprior}
\end{figure}

This LAM-prior distribution can be substituted into the final model for a prior-informed prediction. An example of the LAM-prior distribution can be seen in Fig.~\ref{fig:lam_prior}. The posterior predictive mean with the LAM-prior is given by
\begin{align}
\bm{\mu}\left( y, \eta \mid \mathbf{t} \right) &= \left( \mathbf{K}_{\mathbf{X} \ast} + \bm{\Sigma}_{\mathbf{X}\ast}^{\text{LAM}} \right)^T \left( \mathbf{K}_{\mathbf{XX}} + \bm{\Sigma}_{\mathbf{XX}}^{\text{LAM}} + \bm{\Sigma_s} \right)^{-1} \left( \mathbf{y} + \bm{\mu}_{\ast}^{\text{LAM}} \right), \text{and} \label{eq:pdp_mean} \\
\bm{\Sigma}\left( y, \eta \mid \mathbf{t} \right) &= \left( \mathbf{K}_{\ast\ast} + \bm{\Sigma}_{\ast \ast}^{\text{LAM}} \right) - \left( \mathbf{K}_{\mathbf{X} \ast}^{\text{LAM}} + \bm{\Sigma}_{\mathbf{X}\ast}^{\text{LAM}} \right)^T \left( \mathbf{K}_{\mathbf{XX}} + \bm{\Sigma}_{\mathbf{XX}}^{\text{LAM}} + \bm{\Sigma_s} \right)^{-1} \left( \mathbf{K}_{\mathbf{X}\ast} + \bm{\Sigma}_{\mathbf{X}\ast}^{\text{LAM}} \right). \label{eq:pdp_cov}
\end{align}

\noindent In the equations, $\bm{\mu}\cdot^{\text{LAM}}$ and $\bm{\Sigma}\cdot^{\text{LAM}}$, are computed using the averaging approach defined in Eqs.~\ref{eq:lam_prior_mean} and \ref{eq:lam_prior_cov}. These quantities are evaluated over the latent space and spanwise coordinate $\eta$, either at the training inputs (indicated by subscript $\mathbf{X}$) or at the test inputs (indicated by subscript $\ast$).

Figure~\ref{fig:lam_posterior} presents the posterior predictive distribution of a GP model informed by a LAM-prior. It is evident that the posterior uncertainty is significantly reduced compared to that obtained using a constant mean prior (Fig.~\ref{fig:constMean_posterior}). This reduction is particularly pronounced near the trailing edge, where training data are not present. This improvement arises from two key factors: the variance of the LAM prior is smaller than that of the constant mean prior, and the values and overall trend of the training data are much closer to the mean of the LAM prior than to that of the constant mean prior. 

The physics-driven DKL model, with its predictions informed by prior distributions generated from the LAM, is named the Large Wing Model (LWM). Intuitively, the LWM can be interpreted as a model that learns the three-dimensional correction to the two-dimensional behavior of the underlying airfoil section of the wing. In particular, the model is trained to capture how the spanwise-averaged $C_p$ distribution of a given airfoil varies as a function of the spanwise location and geometric parameters such as aspect ratio, taper ratio, and leading edge sweep. Such an approach yields two main benefits. 

First, the model is data-efficient, achieving good predictive accuracy using relatively little training data. As the aerodynamically feasible $C_p$ behavior is prescribed before training, the model does not need to learn the full mapping from the 64 input variables to the wing's $C_p$ curves, a task that is otherwise extremely high-dimensional. This data efficiency is particularly valuable in wing aerodynamics, where experimental training data for finite wings is significantly more scarce than for airfoils.

Second, the model gains robustness and improved extrapolation capability by leveraging a previously validated two-dimensional aerodynamic foundation. By anchoring its predictions in well-established airfoil behavior, the LWM can generalize more reliably to unseen wing configurations, especially in regions of the input space that are underrepresented in the training data.

\subsection{Model Training}
The LWM was implemented using the open-source Python library, GPyTorch~(Gardner et al., \citeyear{gpytorch}). The training of the model was performed on a single NVIDIA A100 GPU, using the Adam optimizer~(Kingma and Ba,~\citeyear{adam}) to maximize the model's marginal likelihood given the training data. The optimizer had an initial learning rate of $1.0 \times 10^{-3}$, with a step decay of 0.5 every 1000 epochs, resulting in a learning rate of $1.0 \times 10^{-4}$ after 3000 epochs of training.  

The model was trained using exact Gaussian Process (GP) inference, where the covariance matrix of the training set is directly inverted. This approach is typically bottlenecked by high memory demands and computational complexity. However, unlike the LAM, the training data for wing pressure distributions are much more limited in size. As such, the computational burden was significantly reduced, and exact inference remained tractable, delivering reasonable efficiency even without additional approximations or sparse methods commonly employed to scale GP models. These potential venues for improvement is discussed in more detail in Section~\ref{sec:limitations}.

Inspired the approach of Stochastic Weight Averaging (SWA) proposed by Izmailov et al.,~\citeyearpar{izmailov2018}, which equally averages the weights of a neural network traversed by the optimizer, the LWM predictions were generated by restoring the best weights from three different training runs. Similar to its effect on improving the quality of the $C_p$ prediction for the LAM, the SWA procedure was also found to enhance the overall predictive performance of the LWM. 

The mean absolute error in the area enclosed by the predicted $C_p$ curves ($\text{MAE}_{\text{enclosed}}$) was found to be 0.013. This accuracy metric is analogous to an error in $c_l$ and offers an alternative means of evaluating the overall accuracy of the predicted pressure distribution at a given spanwise station. Unlike conventional point-wise metrics such as mean absolute error (MAE) or mean absolute percentage error (MAPE) of $C_p$, $\text{MAE}_{\text{enclosed}}$ captures the integrated discrepancy across the curve.

\subsection{Calculation of Lift and Posterior Constraining via Conditioning}
Given $C_{p,u}\left( \hat{x} \right)$ and $C_{p,l}\left( \hat{x} \right)$, which are the model-predicted pressure coefficients with respect to the $x$-location in conformal coordinates, the sectional lift at a given spanwise location can be approximated with the equation
\begin{align}  
c_l & \approx \frac{1}{2} \cos{\alpha} \int_{-1}^1{\left( C_{p, l}\left(\hat{x} \right) - C_{p, u}\left( \hat{x} \right) \right) d\hat{x}},   \label{eq:cl}  
\end{align}

\noindent under the small angle assumption. Here, the bounds of the integral differ from the typical 0.0--1.0 range due to the change of coordinate systems.

The total lift coefficient can then be calculated by integrating the sectional lift coefficients along the spanwise direction, written as 
\begin{align}  
C_L &= \int_{0}^1{c_l\left(\eta\right) d\eta}. \label{eq:CL}  
\end{align}

As integrals are linear operators, the integral over a Gaussian process preserves Gaussianity. Hence, in the context of aerodynamics, if $C_p$ is modeled as a Gaussian process over the physical domain of chordwise and spanwise directions, the posterior distribution of both $c_l$ and $C_L$ can be described as closed form equations:

\begin{align}
\mu(c_l) &= \left(\int_{c_l} \mathbf{K}_{\ast,X} \ d\hat{x}_* \right)  
\left( \mathbf{K}_{X,X} + \mathbf{\Sigma_s} \right)^{-1} \mathbf{y}, \label{eq:closedform1}\\ 
\sigma^2(c_l) &= \iint_{c_l} \mathbf{K}_{\ast, \ast} \ d\hat{x}_\ast d\hat{x}^\prime_*
- \left(\int_{c_l} \mathbf{K}_{\ast,X} \ d\hat{x}_* \right)  
\left( \mathbf{K}_{X,X} + \mathbf{\Sigma_s} \right)^{-1} 
\left(\int_{c_l} \mathbf{K}_{\ast,X}^T \ d\hat{x}_* \right),  \\ 
\mu(C_L) &= \left(\int_{C_L} \left( \int_{c_l} \mathbf{K}_{\ast,X} \ d\hat{x}_\ast \right) d\eta_\ast \right)  
\left( \mathbf{K}_{X,X} + \mathbf{\Sigma_s} \right)^{-1} \mathbf{y}, \\
\sigma^2(C_L) &= \iint_{C_L}\left( \iint_{c_l} \mathbf{K}_{\ast, \ast} \ d\hat{x}_\ast \ d\hat{x}^\prime_*\right) d\eta_\ast \ d\eta^\prime_\ast \notag\\ 
& \qquad - \left(\int_{C_L} \left( \int_{c_l} \mathbf{K}_{\ast,X} \ d\hat{x}_* \right) d\eta_\ast \right)  
\left( \mathbf{K}_{X,X} + \mathbf{\Sigma_s} \right)^{-1} 
\left(\int_{C_L} \left( \int_{c_l} \mathbf{K}_{\ast,X}^T \ d\hat{x}_* \right) d\eta_\ast \right),  \label{eq:closedform2}
\end{align}

\noindent where the operators $\int_{c_l}$ and $\int_{C_L}$ represent the integral operations required to obtain $c_l$ and $C_L$, respectively. These operations are essentially identical to those in Eqs. \ref{eq:cl} and \ref{eq:CL}, involving integration over the domains of $\hat{x}$ and $\eta$. However, the integrands are the entries within the covariance matrices, rather than the $C_p$ values.

While $c_l$ and $C_L$ are computed from $C_p$ in the LWM, it is important to note that $C_p$ is modeled as a Gaussian process in the latent space, rather than the physical space. The latent variables are obtained through a series of non-linear transformations of the input space via a fully connected neural network. As a result, the mapping between $C_p$ and $\hat{x}$ is non-linear, and Gaussianity is no longer preserved, making the assumptions behind the closed-form equations (Eqs.~\ref{eq:closedform1}--\ref{eq:closedform2}) invalid.

Hence, $C_L$ must be obtained by numerically computing the integral instead, using the Monte Carlo method. A total of 100,000 $C_p$ samples was drawn from the posterior predictive distribution, which ensured convergence of the mean and standard deviation. The integrals, operating directly on the $C_p$ via Eqs.~\ref{eq:cl} and \ref{eq:CL} rather than the covariance matrices, were approximated from these samples. From the resulting data, the mean and standard deviation of $c_l$ and $C_L$ were calculated. This process is described by
\begin{align}  
\mu(c_l) &\approx \frac{1}{n_\text{samples}}\sum_{i=1}^{n_\text{samples}} \frac{1}{2} \cos{\alpha} \int_{-1}^1{\left( C_{p, l, i}\left(\hat{x} \right) - C_{p, u, i}\left( \hat{x} \right) \right) d\hat{x}},\\
\sigma^2(c_l) &\approx \frac{1}{n_\text{samples}-1}\sum_{i=1}^{n_\text{samples}} \left(\frac{1}{2} \cos{\alpha} \int_{-1}^1{\left( C_{p, l, i}\left(\hat{x} \right) - C_{p, u, i}\left( \hat{x} \right) \right) d\hat{x}}-\mu({c_l})\right)^2 \\
\mu(C_L) &\approx \frac{1}{n_\text{samples}}\sum_{i=1}^{n_\text{samples}} \int_{0}^1{c_{l,i}\left(\eta\right) d\eta},  \\
\sigma^2(C_L) &\approx \frac{1}{n_\text{samples}-1}\sum_{i=1}^{n_\text{samples}} \left(\int_{0}^1{c_{l,i}\left(\eta\right) d\eta }-\mu(C_L)\right)^2,
\end{align}

\noindent where $i$ denotes each individual sample and $n_\text{samples}$ is the number of samples, 100,000. 

In many experiments, it is often more practical to measure wing integrated metrics, such as $C_L$, $C_D$, and $C_M$, rather than individual $C_p$ values over the entire surface. This is due to the fact that load cell and balance measurements are typically more accessible than pressure sensors for cost reasons and ease of installation. In these cases, the user may be interested in obtaining the $C_p$ values that resulted in the measured $C_L$, as opposed to using the predicted $C_p$ to calculate the integrated $C_L$.

If probabilistic descriptions of these quantities are known, the information can be used to constrain the posterior distribution, allowing for a potential improvement in $C_p$ predictions and a reduction in variance. Wong et al.~\citeyearpar{wong_linearop} derived analytical expressions for the distribution space conditioned on a linear operator. For the LWM, due to the non-linear element introduced by the neural network, this could not be done analytically. The constraining of the posterior distribution must be performed numerically instead. In this paper, only the effect of conditioning on a known lift coefficient is investigated. However, any integrated metric, namely the aerodynamic force and moment coefficients, can be leveraged. 

A rejection sampling scheme was implemented to draw samples from the posterior predictive distribution that satisfied a prescribed constraint on $C_L$. Samples of $C_p$ were initially drawn from the model, from which corresponding $C_L$ values were computed using Eq.~\ref{eq:CL}. A target distribution $p(C_L) = \mathcal{N}(\mu_{C_{L,c}}, \sigma^2_{C_{L,c}})$ was then used to define an acceptance probability for each sample:
\begin{align}
P_{\text{accept}}(C_L) = \frac{p(C_L)}{M q(C_L)},
\end{align}\label{eq:cond}

\noindent where $q(C_L)$ is the empirical proposal distribution obtained from the initial batch of samples, and $M = \max \left(\frac{p(C_L)}{q(C_L)}\right)$ ensures normalization such that $P_{\text{accept}} \le 1$ for all $C_L$. This yielded a resampled multivariate description of the $C_p$ consistent with the specified distribution of $C_L$.

\begin{figure}[h!]%
\centering
\subfigure[Flowchart of the default model workflow.]{%
\label{fig:flowchart_baseline}%
\includegraphics[width=.99\textwidth]{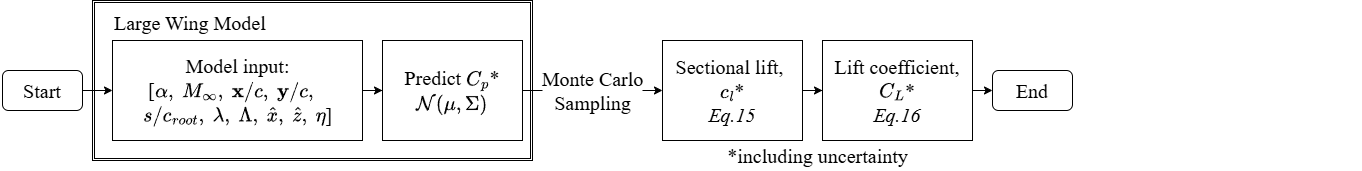}}%
\qquad
\subfigure[Flowchart of the model workflow with $C_L$ conditioning, where the $C_p$ predictions are selectively sampled to satisfy the target $C_L$ distribution.]{%
\label{fig:flowchart_ifCLknown}%
\includegraphics[width=.99\textwidth]{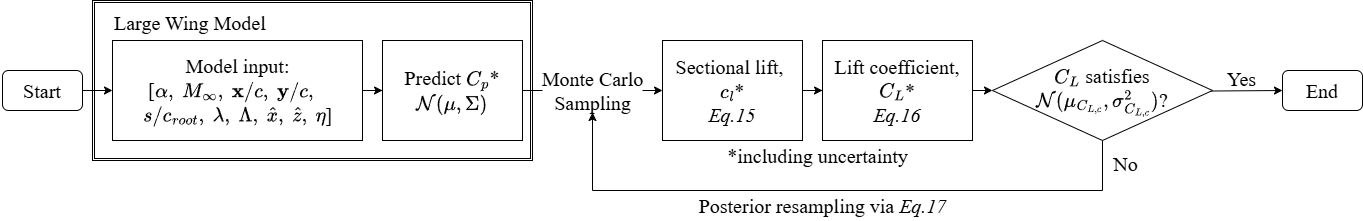}}%
\caption{Flowcharts depicting the Large Wing Model workflow depending on the user input}\label{fig:conditioning_flowchart}
\end{figure}

The resampling process was terminated when the desired number of accepted samples was reached (which defaulted to 5000). Figure~\ref{fig:conditioning_flowchart} illustrates the two flowcharts describing the model workflow: the default workflow without conditioning and a modified workflow which resamples the predictive posterior distribution based on the conditioned $C_L$.

It is important to note that conditioning information can be obtained from various sources, such as load cell and balance measurements. Moreover, the knowledge need not arise solely from experiments; it can also be derived from simulations, low-fidelity models, analytical calculations, or expert knowledge.

\subsection{Model Limitations}\label{sec:limitations}
Since the Large Wing Model's predictions are inherently dependent on those of the Large Airfoil Model (LAM), which serves as the LAM-prior, its limitations, as outlined by Lee et al.,~\citeyear{lee2024_LAM}, also apply here. These include, but are not limited to, difficulties in capturing small geometric variations and perturbations.

Typically, machine learning models produce unreliable predictions when given inputs beyond the range of their training data. However, in this case, the model can remain robust outside its own training range, as long as the inputs fall within the training range of the LAM, which comprises the physics-driven prior. This will be demonstrated in Section~\ref{sec:results}. Users must thus consider the training data range of the LAM, which introduces an additional layer of complexity. In other words, the model's reliability is not solely determined by its own training data, making its behavior less intuitive.

Another limitation of the model lies in its scalability, which stems from the inherent limitations of Gaussian Processes (GPs). The GP component of the LWM requires storing the full $13,132 \times 13,132$ covariance matrix during both training and prediction. While this is less restrictive than in the LAM due to the smaller number of data points, it nonetheless remains essential to ensure that the model can accommodate the continuous expansion of experimental data. The improvement in scalability could be achieved via variational methods such as Stochastic Variational Deep Kernel Learning (SV-DKL by Wilson et al.,~\citeyear{svdkl}). This framework significantly improves memory requirements and computational efficiency by enabling mini-batch training (loading only a subset of data at a time), stochastic variational inference (Hensman et al., \citeyear{hensman2014}), and sparse Gaussian Processes (Titsias, \citeyear{titsias09}). These techniques reduces peak memory usage, making it feasible to train larger models. 

Currently, the model predictions are made under the assumption that the airfoil section remains uniform across the entire span of the wing. In practical aerodynamic design, however, airfoil geometry often varies along the span to satisfy specific performance or structural requirements. Accommodating such spanwise variation would likely necessitate a reparameterization of the training data to incorporate spatially dependent airfoil characteristics, thereby enabling the model to generalize to wing configurations more representative of current designs. To address the limitation, the model could be extended by introducing the spanwise coordinate as an additional input feature, thereby allowing it to condition its predictions on local geometry. This approach would require the training data set to incorporate airfoil shapes sampled at various spanwise locations, each annotated with the corresponding spanwise position.

Lastly, the model does not incorporate any explicit treatment of the wing tip and implicitly assumes a rounded tip geometry for all wings, as most experimental test articles have rounded tip geometry. However, prior studies (e.g., McAlister and Takahashi,~\citeyear{nasa3151}) have shown that sharp-edged wing tips, in the absence of any tip treatment, can exhibit significantly different tip vortex behavior, which in turn affects the pressure distribution near the wing tip. Currently, the model is unable to capture such effects, and there is a lack of comprehensive data sets in the literature that systematically investigate sharp wing tips across a range of wing configurations. As such, this limitation must be taken into account when interpreting the model’s predictions.

\section{Results and Discussion}\label{sec:results}
To evaluate the model's predictive performance and its applicability to wings of airfoils that the model has not seen, all data from NACA 0015 wings were excluded from the training set. The test cases were strategically selected to represent various combinations of airfoil sections, planform geometries, and operating conditions, as outlined in Table~\ref{table:testcase}.

\begin{table}
\centering
\renewcommand{\arraystretch}{1.2}
\begin{tabular}{ |c||c|c|c|c|c|c| }
 \hline
 Case & Airfoil & $\alpha$ [deg] & $M_\infty$ [1] & $c_\text{root}$ [m] & $s/c_\text{root}$ [1] & Source \\
 \hline
 1 & NACA 0012 & $8.85$ & 0.13 & 1.0 & 2.95 & NASA TN D-8307~\citeyearpar{nasa8307} \\
 2 & NACA 0012 & $2.01$ & 0.60 & 0.406 & 2.0 & NASA TM 104211~\citeyearpar{nasa104211}\\ 
 3 & NACA 0015 & $4.00$ & 0.17 & 0.518 & 3.3 & NASA TP 3151~\citeyearpar{nasa3151} \\
 \hline
\end{tabular}
\caption{Wing geometry, operating conditions, and source experiments of the test cases.}
\label{table:testcase}
\end{table}

\begin{itemize}
    \item \textbf{Test case 1}: Involves a wing with a NACA 0012 section. Both the airfoil and planform geometry were present in the training data, with only the operating conditions ($\alpha$ and $M_\infty$) modified. This represents a purely interpolative task, which machine learning models typically handle well.
    \item \textbf{Test case 2}: Introduces a different aspect ratio and a higher freestream Mach number ($M_\infty$), pushing into the compressible flow regime. Although the flow conditions differ significantly, the airfoil section remains NACA 0012, which the model has seen during training.
    \item \textbf{Test case 3}: Features a NACA 0015 airfoil with a different aspect ratio. Since the NACA 0015 section was entirely excluded from training, this case poses a strong test of the model's predictive ability, especially challenging given the relatively sparse data set.
\end{itemize}
All test cases were limited to rectangular wings to isolate the effects of airfoil shape and flow regime.

\subsection{Grid Flexibility}\label{subsec:gridflex}
As a Gaussian Process-based model, the LWM support continuous, non-parametric regression. Once trained, the model can generate predictions of $C_p$ at arbitrary query locations, regardless of the original training grid. This grid flexibility serves as an advantage over neural network-based ML models which typically produce outputs on pre-defined grids. The predictions can be as coarse or fine as the user desires. However, unlike CFD, the results are completely independent of the grid resolution.

This property can be utilized in various ways. For instance, the  model's predictions can be formatted in such a way that it mimics that of a grid defined by a CFD mesh. This enables compatibility with existing visualization and analysis tools, enhancing not only data visualization efficiency but also the comparability between CFD and machine learning results. 

\begin{figure}[h!]
\centering
\includegraphics[width=.5\textwidth, trim=0 300 0 0, clip]{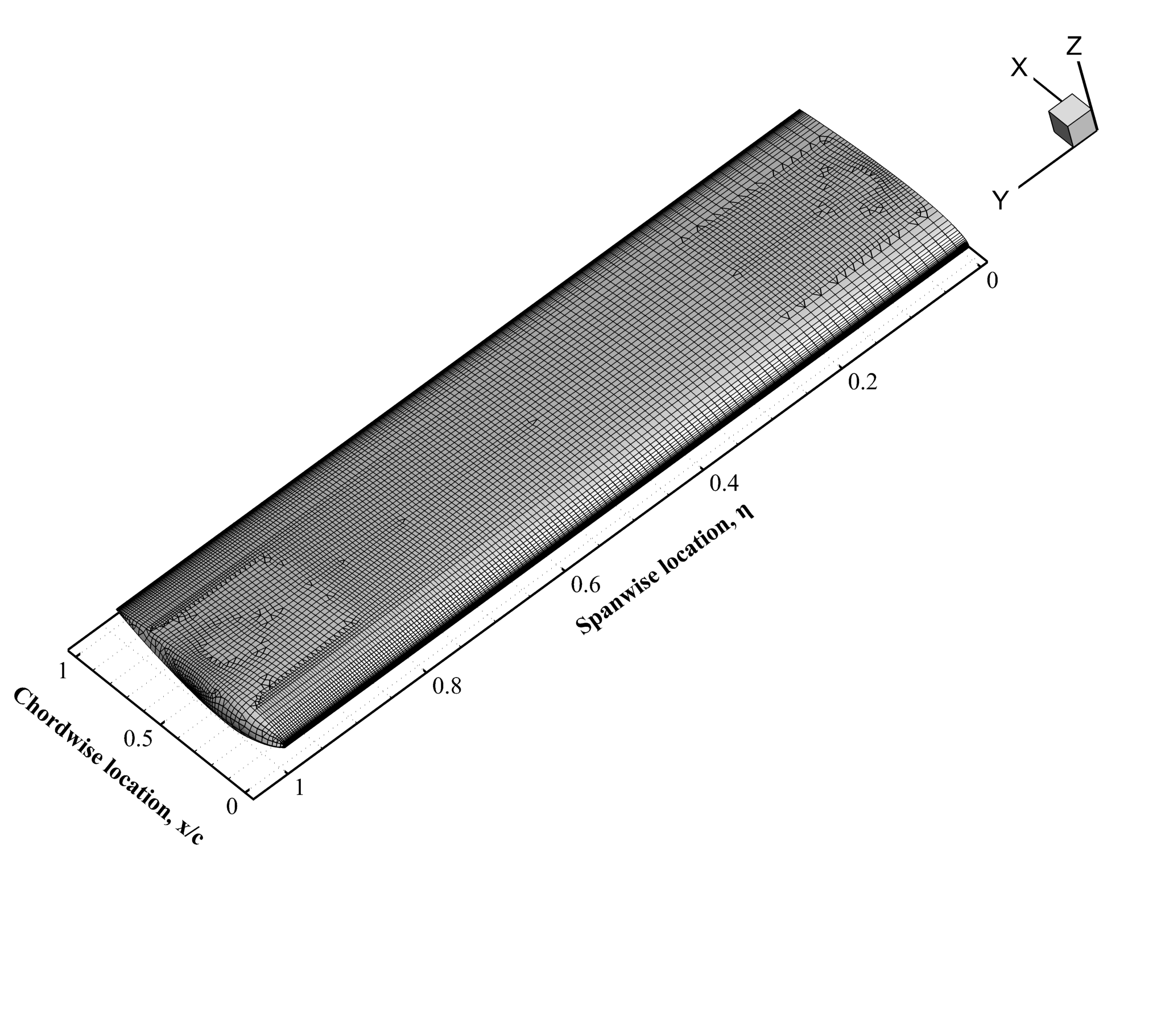}
\caption{Surface mesh node coordinates used for prediction. The node coordinates on the upper and lower surfaces are identical}\label{fig:prediction_grid}
\end{figure}

In this work, the predicted results are presented using this feature, by evaluating predictions directly on an existing CFD mesh. Specifically, a reference mesh of a generic rectangular wing was used, from which only the surface nodes were extracted---resulting in a total of 25,475 points comprising both triangular and quadrilateral finite elements. Figure~\ref{fig:prediction_grid} shows the configuration of this surface mesh. By generating predictions from the LWM model at these
node coordinates, visualization of contour plots across the wing could be performed. This reference mesh was used consistently across all test cases.

\subsection{Test Case 1: NACA 0012 Wing, Incompressible Regime}
\begin{figure}[h!]
     \centering
     \subfigure[Upper surface]{\includegraphics[width=0.99\textwidth]{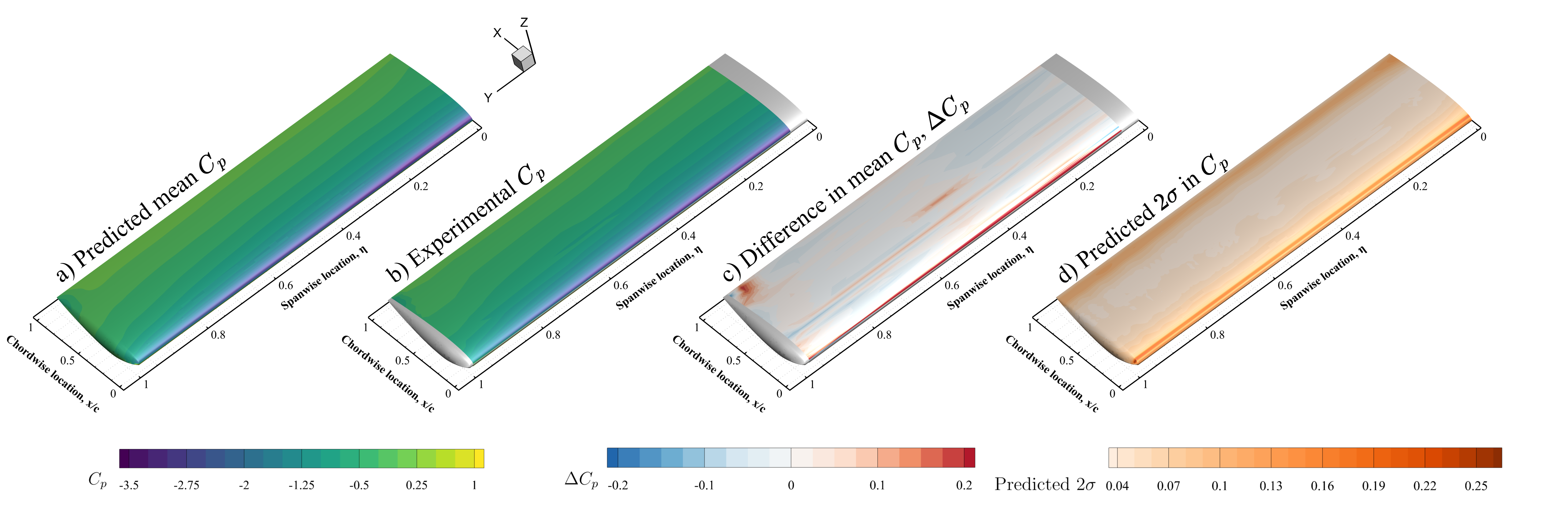}} 
     \subfigure[Lower surface]{\includegraphics[width=0.99\textwidth]{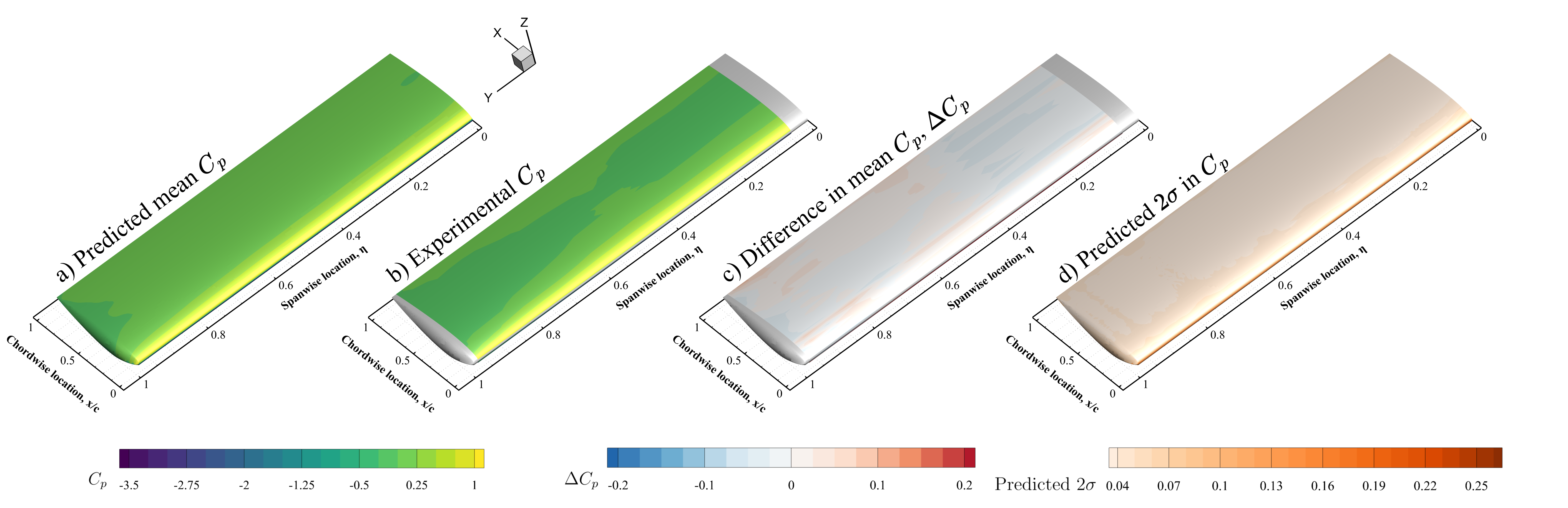}}
     \caption{Test case 1: contour plots of the predicted mean $C_p$, experimentally measured $C_p$~(Yip and Shubert,~\citeyear{nasa8307}), the difference between the predicted mean and the experiment, and the predicted two standard deviations over the wing}
     \label{fig:TC1_contour}
\end{figure}

Figure~\ref{fig:TC1_contour} depicts the LWM results for test case 1: the contours of the mean predicted wing $C_p$ distribution, the corresponding experimental measurements, the error between the predicted mean and experimental data, and the predicted uncertainty (95\% confidence interval, $2\sigma$). The predictions were found to yield good agreement for both surfaces, with the mean absolute error (MAE) in $C_p$ and the mean absolute error in enclosed area ($\text{MAE}_\text{enclosed}$) of 0.044 and 0.014, respectively. It is notable that the predicted mean $C_p$ distribution is much smoother than that of the experimental measurements, which results in jaggedness in the contour plot. This is attributed to the fact that the Gaussian process layer of the LWM provides a ``noise-free'' prediction of the surface pressures. Consequently, bands in the error contours can be observed, corresponding to instances where measurement noise are present in the original measurements.

The experimental contour plot shows that the tip vortex induces a distortion in the trailing edge $C_p$ on the upper surface at approximately $\eta \ge 0.970$. A similar distortion is observed in the predicted $C_p$, demonstrating the model’s ability to capture three-dimensional effects caused by the rotational flow of the tip vortex. This capability is particularly noteworthy, as such $C_p$ fluctuations cannot be captured by lower-fidelity solvers, such as vortex lattice methods, which neglect viscous effects and offer limited wake modeling. However, a relatively large error ($\Delta C_p \approx 0.21$) is observed near the wing tip, due to a slight mismatch in the affected region. Specifically, the model predicted an earlier onset of tip vortex-induced pressure distortion compared to the experimental measurements. 

The high errors found at the leading edge can be attributed to the overestimation of the suction for certain regions and the sensitivity of predictions to the leading edge pressure gradient. A minor under- or over-prediction of the leading edge pressure gradient can result in an exaggeration of the error at individual points. This issue is particularly pronounced in LWM, as it is trained on manually digitized data, where each chordwise data point carries some level of uncertainty. Consequently, reporting the error between predicted and measured $C_L$, along with the mean absolute error (MAE) in $C_p$, provides a more intuitive metric for assessing the accuracy of the $C_p$ curve shape. For test case 1, the model predicted a $C_L$ of $0.659 \pm 0.005$, compared to the known value of $0.667$. This corresponded to $|\Delta{C_L}| = -0.008$, or $-1.259\%$.

As for the 95\% confidence interval contours of $C_p$, the regions of greatest uncertainty are located at the leading and trailing edges on the upper surface. The large $2\sigma$ values at the leading edge are likely due to spanwise variations in the suction peak magnitude present in the training data, which arise from three-dimensional effects of finite wings. The increased uncertainty at the trailing edge can likely be attributed to the absence of pressure measurements in many experiments at that location. Additionally, the trailing edge, particularly in the outboard region, experiences considerable variation due to the impingement of the tip vortex. Nonetheless, while these regions exhibit the highest uncertainty within the domain, the relative magnitude of the uncertainty remains small compared to the $C_p$ values themselves, and thus does not pose a significant concern.

\begin{figure}[h!]
\centering
\includegraphics[width=.85\textwidth]{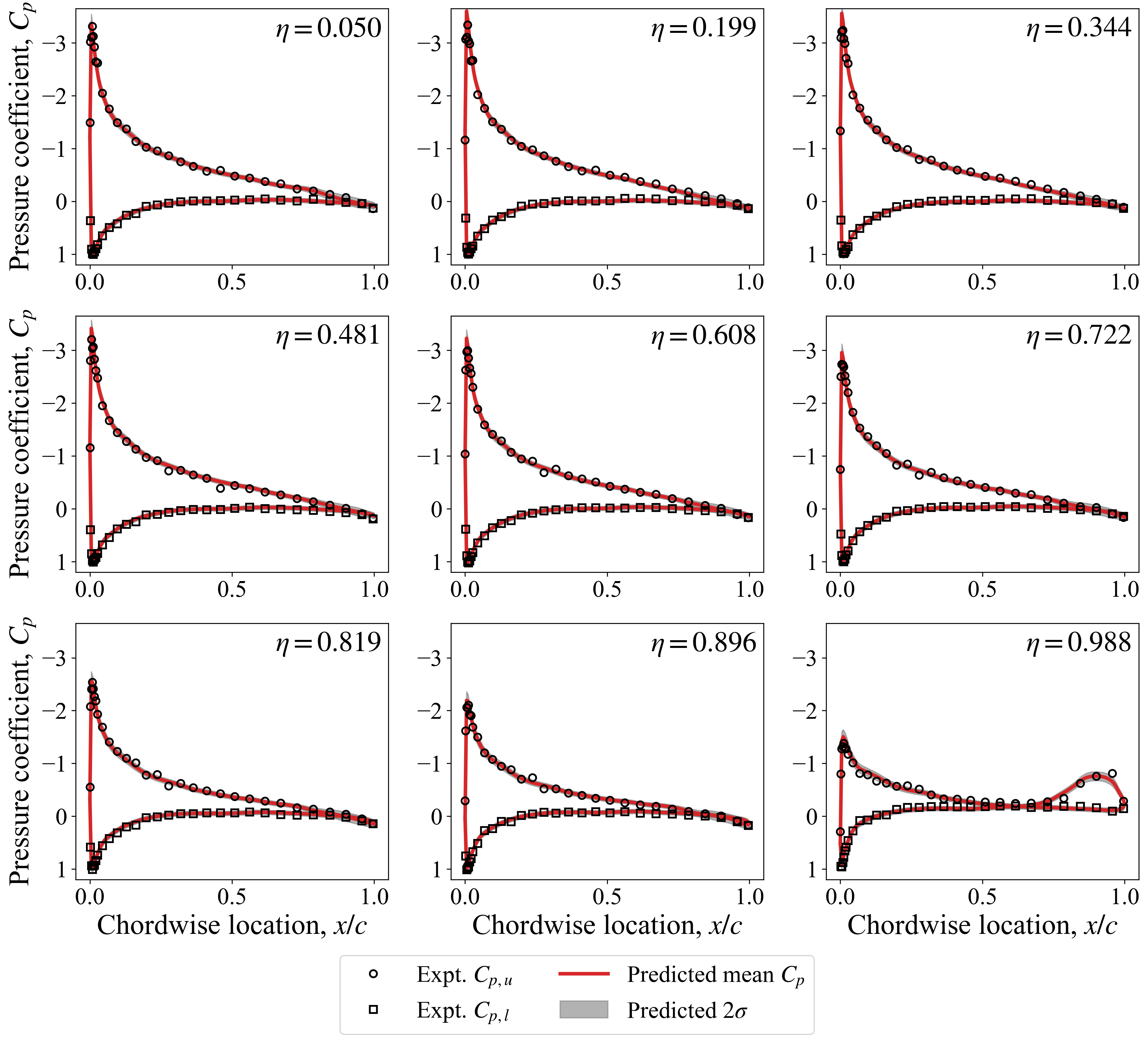}
\caption{Test case 1: Comparison between measured and predicted chordwise $C_p$ distributions at multiple spanwise stations. The measurements are from Yip and Shubert,~\citeyear{nasa8307}}
\label{fig:TC1_slice}
\end{figure}

The chordwise $C_p$ distributions presented in Fig.~\ref{fig:TC1_slice} provide additional insight into how the experimental data compare against the model predictions. Overall, the experimental measurements and the predictions correlate very well, although the suction peak magnitude at the $0.199 \le \eta \le 0.481$ range is overestimated. As also observed in the contour plots (Fig.~\ref{fig:TC1_contour}), differences are observed at the trailing edge for $\eta = 0.988$. The maximum magnitude of the tip vortex effect at this spanwise station was underpredicted. 

\subsection{Test Case 2: NACA 0012 Wing, Compressible Regime}
\begin{figure}[h!]
\centering
\includegraphics[width=.65\textwidth]{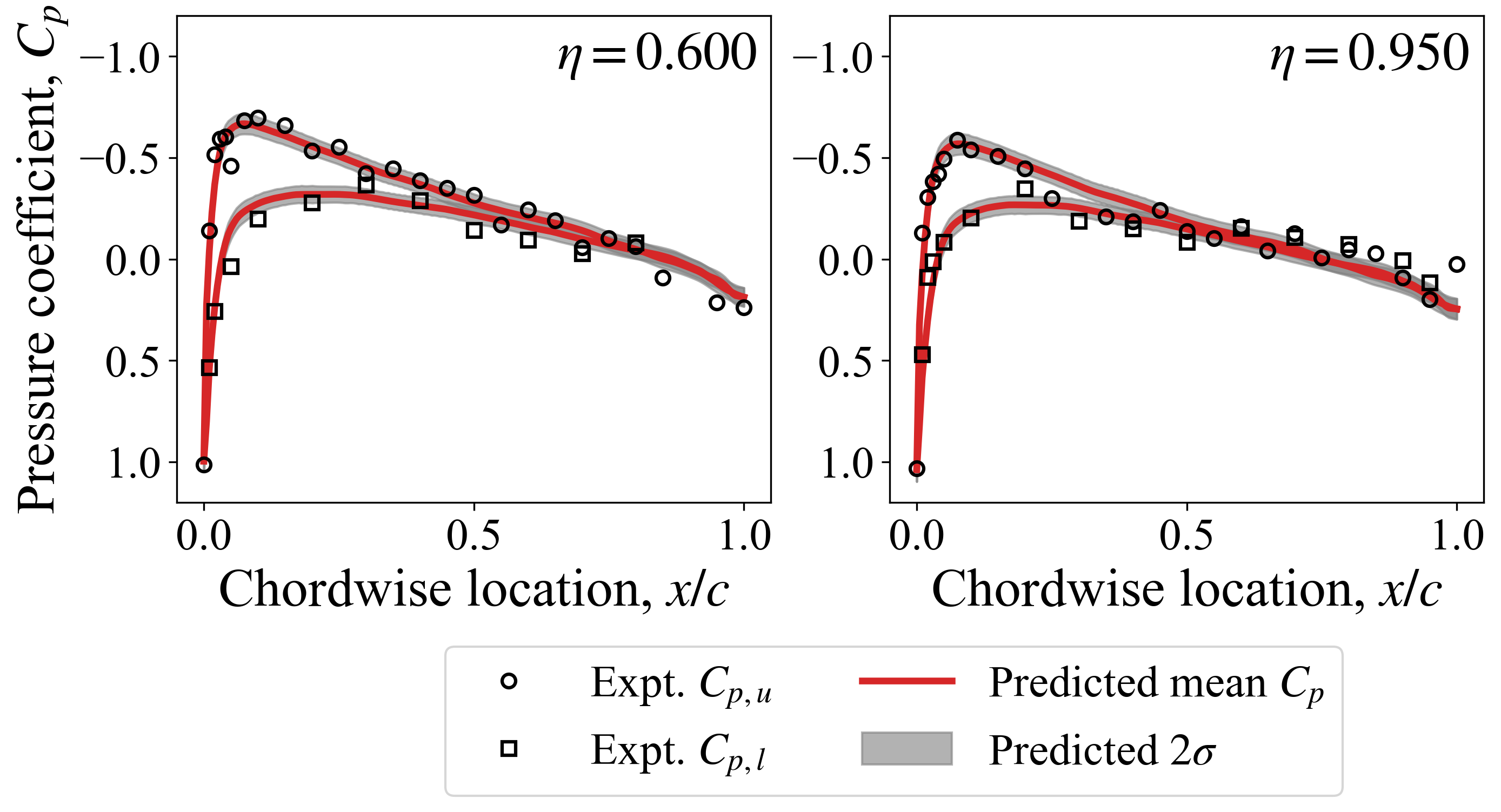}
\caption{Test case 2: Comparison between measured and predicted chordwise $C_p$ distributions at multiple spanwise stations. The measurements are from Rivera et al.,~\citeyear{nasa104211}}
\label{fig:TC2_slice}
\end{figure}

Test case 2 is characterized by a decrease in the aspect ratio and a significant increase in the freestream Mach number, marking a transition into the compressible flow regime. These modifications to the geometry and the operating conditions alter the wing $C_p$ characteristics, which the model must accurately predict. Figure~\ref{fig:TC2_slice} captures the chordwise distribution of $C_p$ at different spanwise slices. As measurements were only available at two locations ($\eta = 0.600, 0.950$), the contour plot is not presented for this test case. 

Although the experimental measurements exhibit significant inherent noise, the model’s predictions still show strong correlation with the data. At higher $M_\infty$, the pressure difference between the upper and lower surfaces becomes more pronounced compared to the incompressible regime, and the model accurately captures these compressibility effects. Quantitatively, the MAE in $C_p$ across the two sections was calculated to be 0.062, while the $\text{MAE}_\text{enclosed}$ was 0.029. These error magnitudes are slightly higher than those in test case 1. However, given that the ground truth data in this case are substantially noisier than in test case 1, the performance can still considered acceptable. The predicted $C_L$ was $0.126 \pm 0.002$, although no reference value was provided in the source material for comparison.

\subsection{Test Case 3: NACA 0015 Wing}
Before investigating the model accuracy for a NACA 0015 wing, it is useful to elucidate the contribution of the LAM-prior in predicting $C_p$ distributions for airfoil sections that were not part of the training data. This evaluation directly relates to the model's extrapolation capability. 

\begin{figure}[h!]%
\centering
\subfigure[NACA 0012 section (included in the training data)]{%
\label{fig:lamp_seenAF}%
\includegraphics[width=.35\textwidth]{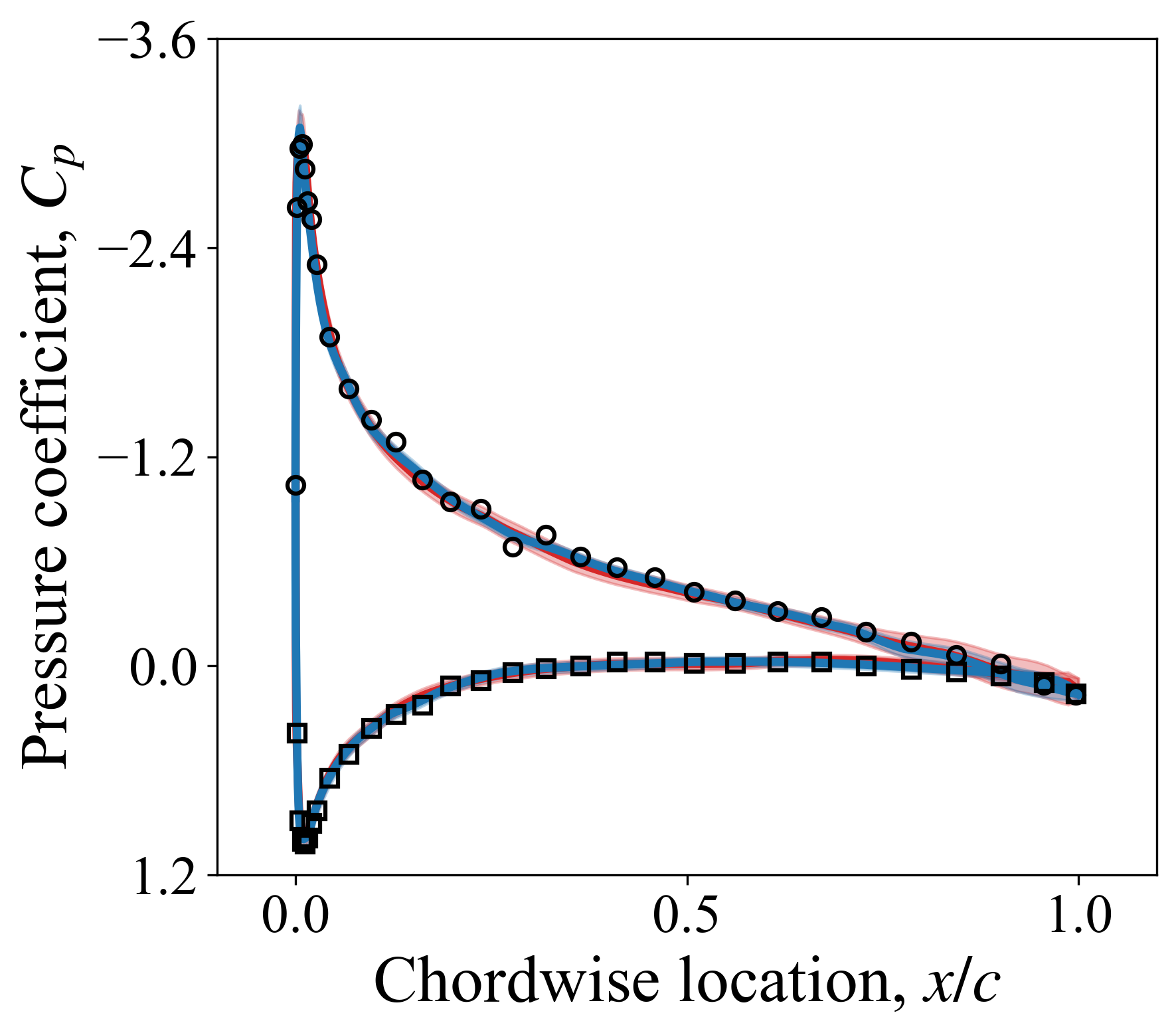}}%
\qquad
\subfigure[NACA 0015 section (not included in the training data)]{%
\label{fig:lamp_unseenAF}%
\includegraphics[width=.35\textwidth]{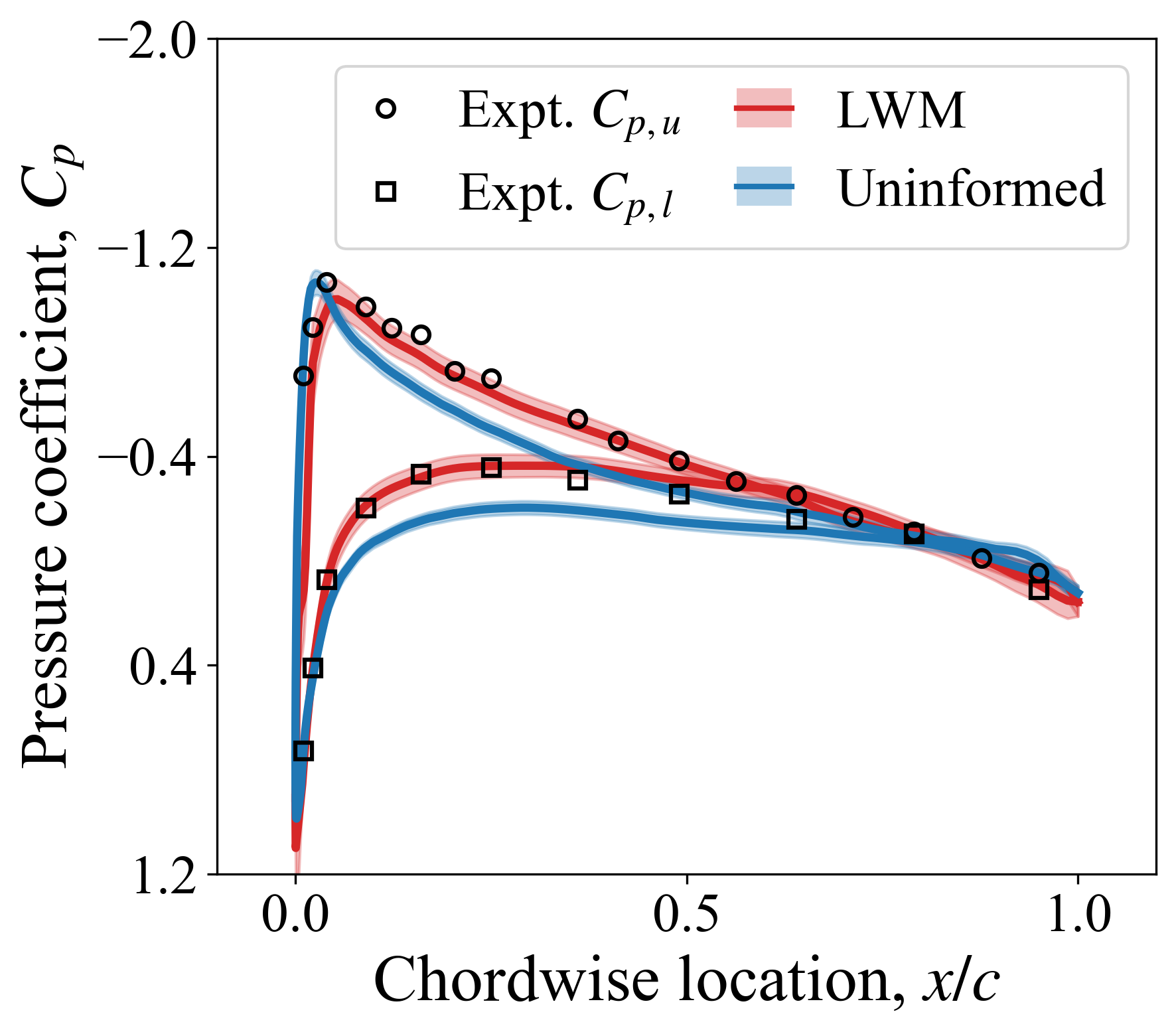}}%
\caption{Comparison between an uninformed baseline model and the physics-driven Large Wing Model for two wings with different airfoil sections}\label{fig:lamp_effect}
\end{figure}

Figure~\ref{fig:lamp_effect} compares the predictions of an uninformed (no LAM-prior) baseline model and the LWM for two wings with different airfoil sections: NACA 0012 and NACA 0015. Both models share the same DKL architecture. The lack of an informed prior does not detrimentally affect the uninformed model performance for a wing with a NACA 0012 section, the data of which are abundantly available within the training data set. This is evidenced by Fig.~\ref{fig:lamp_seenAF}, where the uninformed model prediction is essentially identical to that of the LWM. However, for prediction on a NACA 0015 wing (an airfoil section deliberately removed from the training set), the accuracy of the uninformed model decreased drastically as seen in Fig.~\ref{fig:lamp_unseenAF}. This is due to the limited diversity of airfoil sections in the finite wing experiments used for training, which prevents the baseline model from generalizing effectively. In contrast, the LWM's predictions are based on the LAM-predicted $C_p$ of the NACA 0015 airfoil. The findings corroborate the claim that this paper's physics-driven approach substantially enhances the prediction accuracy compared to that of an uninformed model.

It should be noted that the measurements of test case 3 were corrected for tunnel blockage effects based on the information provided in the source material, resulting in slight differences from the original figures. This adjustment was made following the recommendations of the authors of the original paper, McAlister and Takashi~\citeyearpar{nasa3151}. 

\begin{figure}[h!]
     \centering
     \subfigure[Upper surface]{\includegraphics[width=0.95\textwidth]{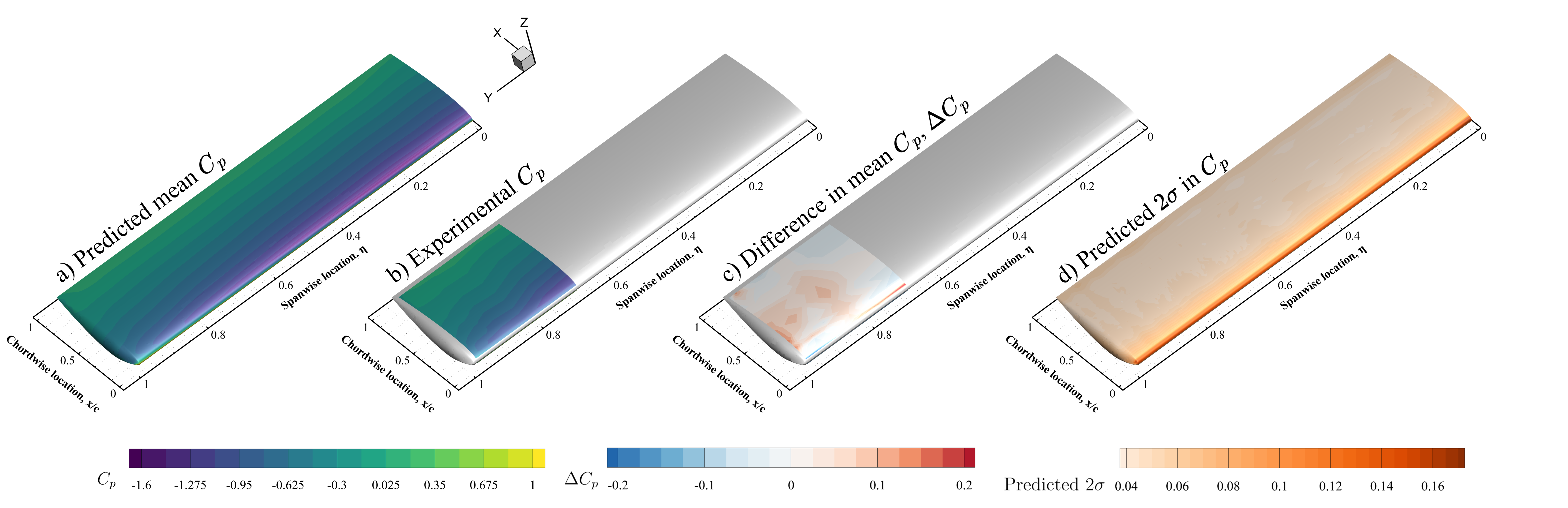}} 
     \subfigure[Lower surface]{\includegraphics[width=0.95\textwidth]{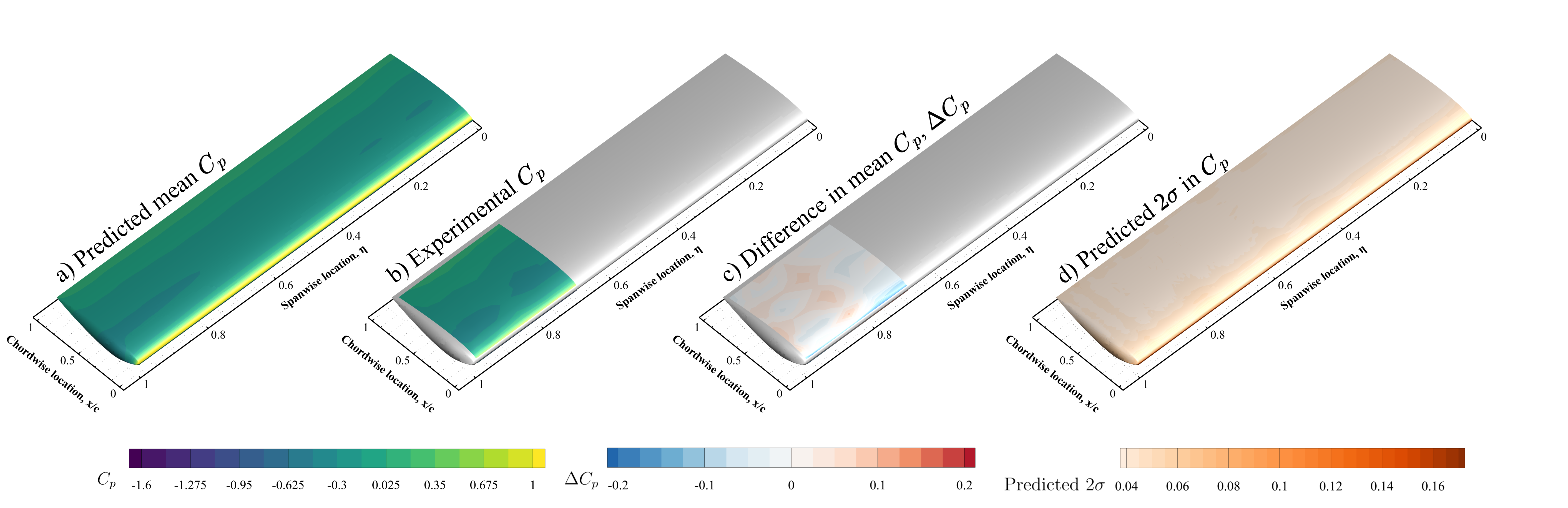}}
     \caption{Test case 3: Contour plots of the predicted mean $C_p$, experimentally measured $C_p$~(McAlister and Takahashi~\citeyear{nasa3151}), the difference between the predicted mean and the experiment, and the predicted two standard deviations over the wing. Note that the experimental results were only for the outboard region}
     \label{fig:TC3_contour}
\end{figure}

The contour plots for test case 3 (Fig.~\ref{fig:TC3_contour}) revealed similar trends to those of test case 1, with strong qualitative agreement between the experimental measurements and model predictions. This is a noteworthy result, as it represents an extrapolative prediction for an airfoil section that was not seen during training. Notably, the model successfully predicts the pressure undulation at the wing tip upper surface. The ability to predict wing tip effects across different airfoil sections shows that the model was trained to generalize effectively. As with test case 1, the greatest difference in $C_p$ occurred at the leading edge suction peak and the trailing edge of the wing tip region. These are also the regions with the highest predicted $2\sigma$ in $C_p$.

\begin{figure}[h!]
\centering 
\includegraphics[width=.80\textwidth]{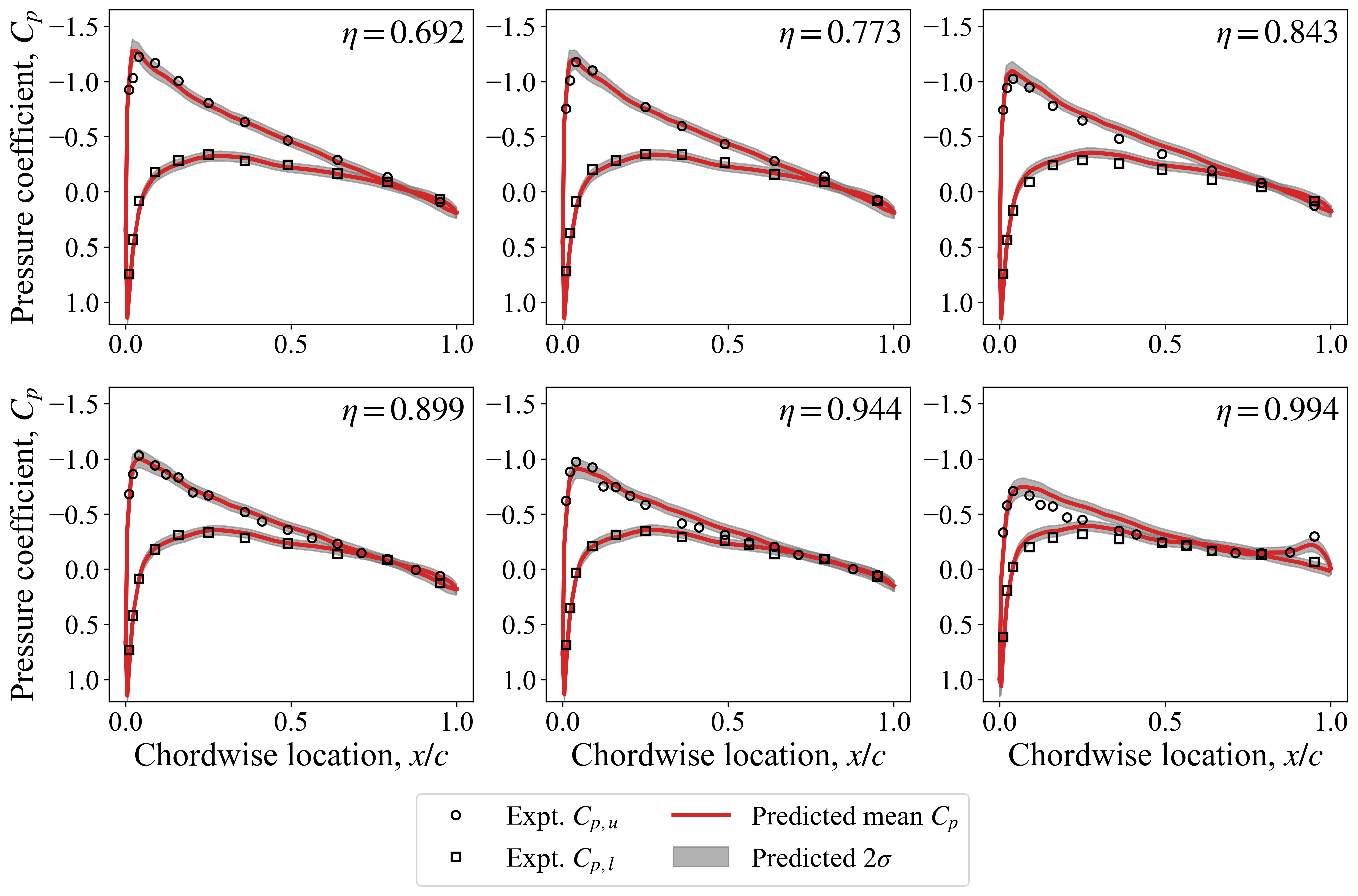}
\caption{Test case 3: comparison between measured and predicted chordwise $C_p$ distributions at multiple spanwise stations. The measurements are from McAlister and Takahashi,~\citeyear{nasa3151}}
\label{fig:TC3_slice}
\end{figure}

Figure~\ref{fig:TC3_slice} provides additional insight into the observed discrepancy. The chordwise $C_p$ distributions reveal a good correlation between the predictions and the measurements. However, the model yields a sharper suction peak for $\eta \le 0.773$ range, causing the larger $\Delta C_p$ observed in Fig.~\ref{fig:TC3_contour}. An overprediction of the $C_p$ magnitude at the wing tip of $\eta=0.994$ can also be observed, which would lead to some discrepancy in the sectional lift coefficient, $c_l$, at the wing tip. This contributed to the slight overprediction of the overall $C_L$, calculated to be $0.359 \pm 0.003$. The value corresponds to a 1.700\% error compared to the measured value of 0.353. The MAE of $C_p$ and $\text{MAE}_\text{enclosed}$ were calculated to be 0.052 and 0.016, respectively. 

\subsection{Effect of Spatially Varying Hyperparameters}
It was previously argued that the spatially varying hyperparameters, where the kernel length scale is modified with respect to the spanwise location, $\eta$, would be beneficial for the model predictive performance. In this section, this effect is analyzed in greater detail. Test case 1 is selected for this analysis, as it offers the most comprehensive spanwise coverage, allowing for direct comparisons between the predicted results and experimental measurements at both the wing root and tip.

\begin{figure}[h!]
\centering
\includegraphics[width=.95\textwidth]{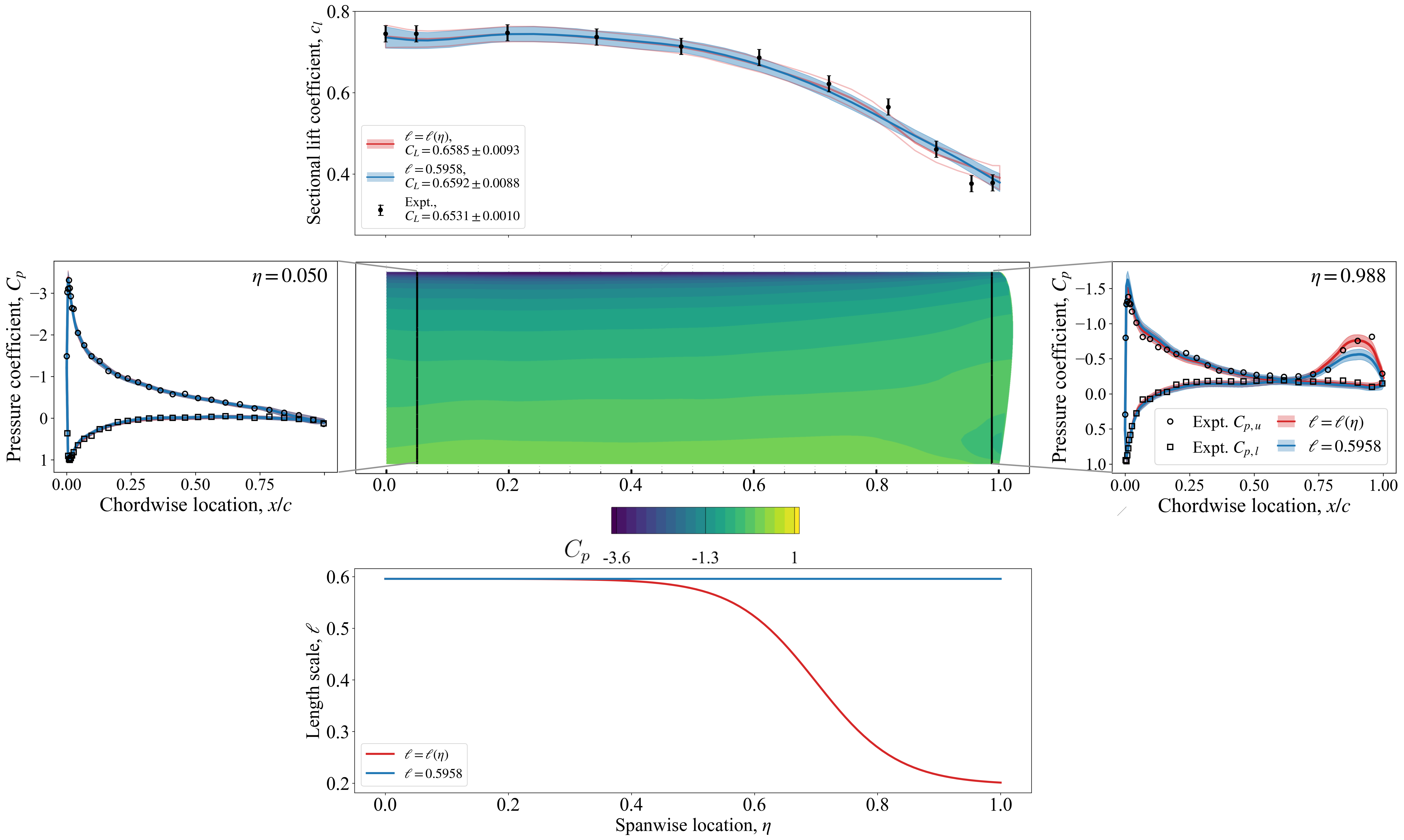}
\caption{Comparison of model behavior between constant and spatially varying spanwise length scales for test case 1. (Top) Predicted spanwise $c_l$ distributions. (Middle) Mean $C_p$ contour for a constant value of $\ell=0.5958$. (Left) Measured vs. predicted chordwise $C_p$ at $\eta=0.05$. (Right) Measured vs. predicted chordwise $C_p$ for $\eta=0.988$. (Bottom) Spatially uniform vs. varying length scales}
\label{fig:hiEll}
\end{figure}

Figure~\ref{fig:hiEll} presents several results obtained when the model employs a single length scale value ($\ell = 0.5958$), the upper bound of the two values derived from the optimization of the spatially varying hyperparameters. This fixed length scale is visualized in the bottom subfigure. The middle contour plot reveals that the transition in the $C_p$ distribution is overly smooth, leading to a slow and gradual variation in the spanwise direction. This smoothness hinders the model’s ability to accurately capture the localized effects of the tip vortex. Notably, the tip vortex-affected region near the wing tip appears excessively large in the visualization, which leads to an overprediction of $c_l$, as shown in the top subfigure.

Additionally, the chordwise $C_p$ profile at $\eta = 0.988$ exhibits significant discrepancies in magnitude. As discussed in Section~\ref{subsec:svh}, this stems from the fact that larger length scales tend to emphasize broader spatial trends while suppressing localized fluctuations. In summary, relying solely on a large length scale introduces limitations in capturing small-scale aerodynamic features.

\begin{figure}[h!]
\centering
\includegraphics[width=.95\textwidth]{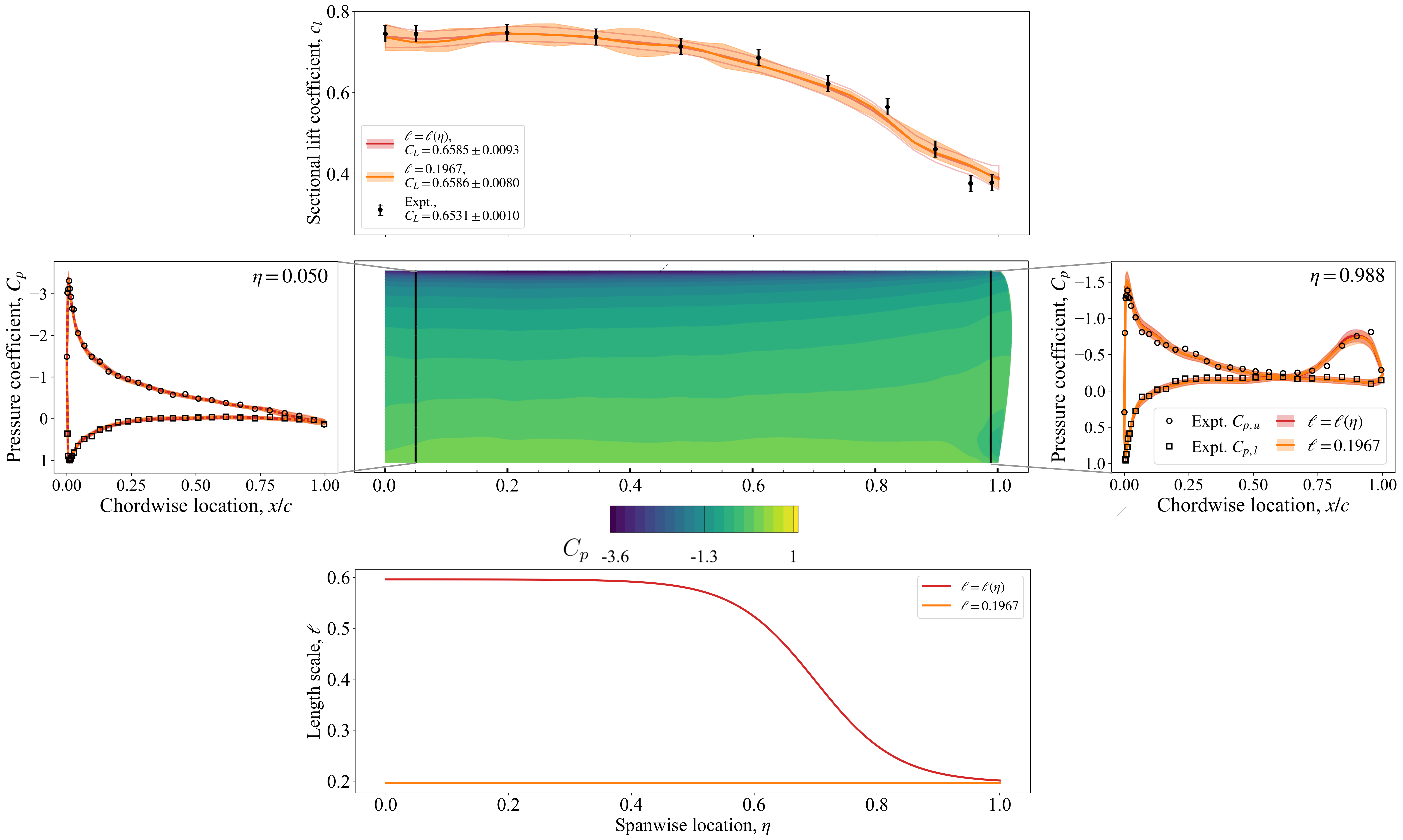}
\caption{Comparison of model behavior between constant and spatially varying spanwise length scales for test case 1. (Top) Predicted spanwise $c_l$ distributions. (Middle) Mean $C_p$ contour for a constant value of $\ell=0.1967$. (Left) Measured vs. predicted chordwise $C_p$ at $\eta=0.05$. (Right) Measured vs. predicted chordwise $C_p$ for $\eta=0.988$. (Bottom) Spatially uniform vs. varying length scales}
\label{fig:loEll}
\end{figure}

On the other hand, results presented in Fig.~\ref{fig:loEll} utilize the lower bound of the spatially varying hyperparameters ($\ell=0.1967$). This is portrayed in the bottom subplot. The middle contour plot and the chordwise $C_p$ slice at the wing tip indicate that the extent of the tip vortex-affected region is more reasonable, and the predicted magnitude of $C_p$ is also in better agreement with the reference data. This comes at a cost of the smoothness of the prediction, with minor fluctuations of $C_p$ in the spanwise direction being observed. The oscillatory behavior is further corroborated by the sectional lift coefficient distribution plot, as the $c_l$ shows an oscillatory behavior in both the mean and the 2$\sigma$ uncertainty range compared to the original LWM with spatially varying hyperparameters.

In conclusion, accurately predicting tip vortex-induced distortions in the wing tip $C_p$ necessitates a smaller kernel length scale. In contrast, inboard regions, where such effects are minimal, benefit from a larger length scale that promotes smoother transitions in $C_p$ along the span. The spatially varying length scale approach introduced in this work effectively achieves both objectives, combining localized sensitivity with global smoothness to deliver improved predictive performance.

\subsection{Posterior Constraining Based on Measured Lift}
The LWM was found to yield strong accuracy across all three cases. However, for test case 3, small but consistent differences in $C_p$ magnitude could be observed. For an integrated metric such as lift, these consistent errors can accumulate over the entire integral domain. This effect is especially pronounced for a wing, as errors can propagate in both chordwise and spanwise directions.

\begin{figure}[h!]
\centering
\includegraphics[width=.6\textwidth]{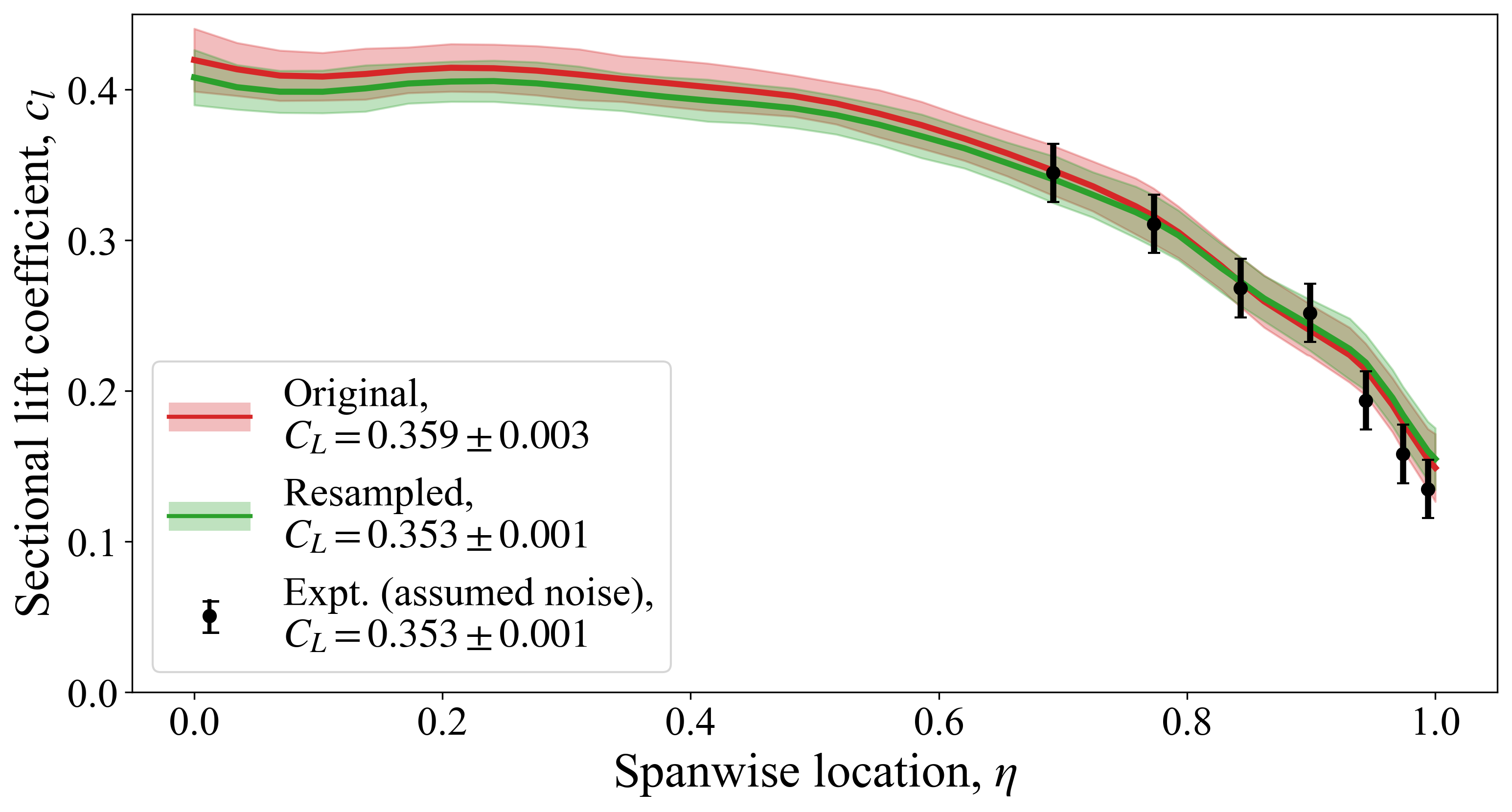}
\caption{Spanwise $c_l$ distribution of the original predictive posterior distribution and the resampled predictive posterior distribution via conditioning on the known $C_L$. Presented results are for test case 3}
\label{fig:cond_spanwiselift}
\end{figure}

The predicted spanwise $c_l$ distribution for test case 3 is presented as a red line in Fig.~\ref{fig:cond_spanwiselift}. Notably, the sectional lift coefficient at the wing tip was non-zero due to the tip vortex effects. This contrasts with linearized potential flow-based methods. For example, a vortex lattice method code that models the wing as a lattice of horseshoe vortices would indicate $c_l(\eta =1.0) = 0$. This would yield significant errors relative to the experimental data. 

Figure~\ref{fig:cond_spanwiselift} also illustrates that the sectional lift is mostly overpredicted. The maximum value of this error was found to be 0.021, resulting in an overprediction of the total lift coefficient by 1.7\%, as noted previously. The model was thus conditioned on the known $C_L$, provided in the original experimental document, to investigate how the posterior predictive distribution would be affected. The $2\sigma$ range for $C_L$ was set to 0.001.

With the conditioning, the agreement between the experiment and the prediction of the sectional lift distribution (green line) improved especially in the non-tip region ($\eta \le 0.90$). While there remains some overprediction of $c_l$ at the very wing tip, all experimental measurements fell within the 95\% confidence bound of the predictions. $C_L$ had perfect agreement as the predictive distribution had been conditioned on the value. By constraining the variance of the overall $C_L$ to a comparatively smaller number, the confidence interval ranges at all spanwise locations were reduced slightly.

\begin{figure}[h!]
\centering
\includegraphics[width=.4\textwidth]{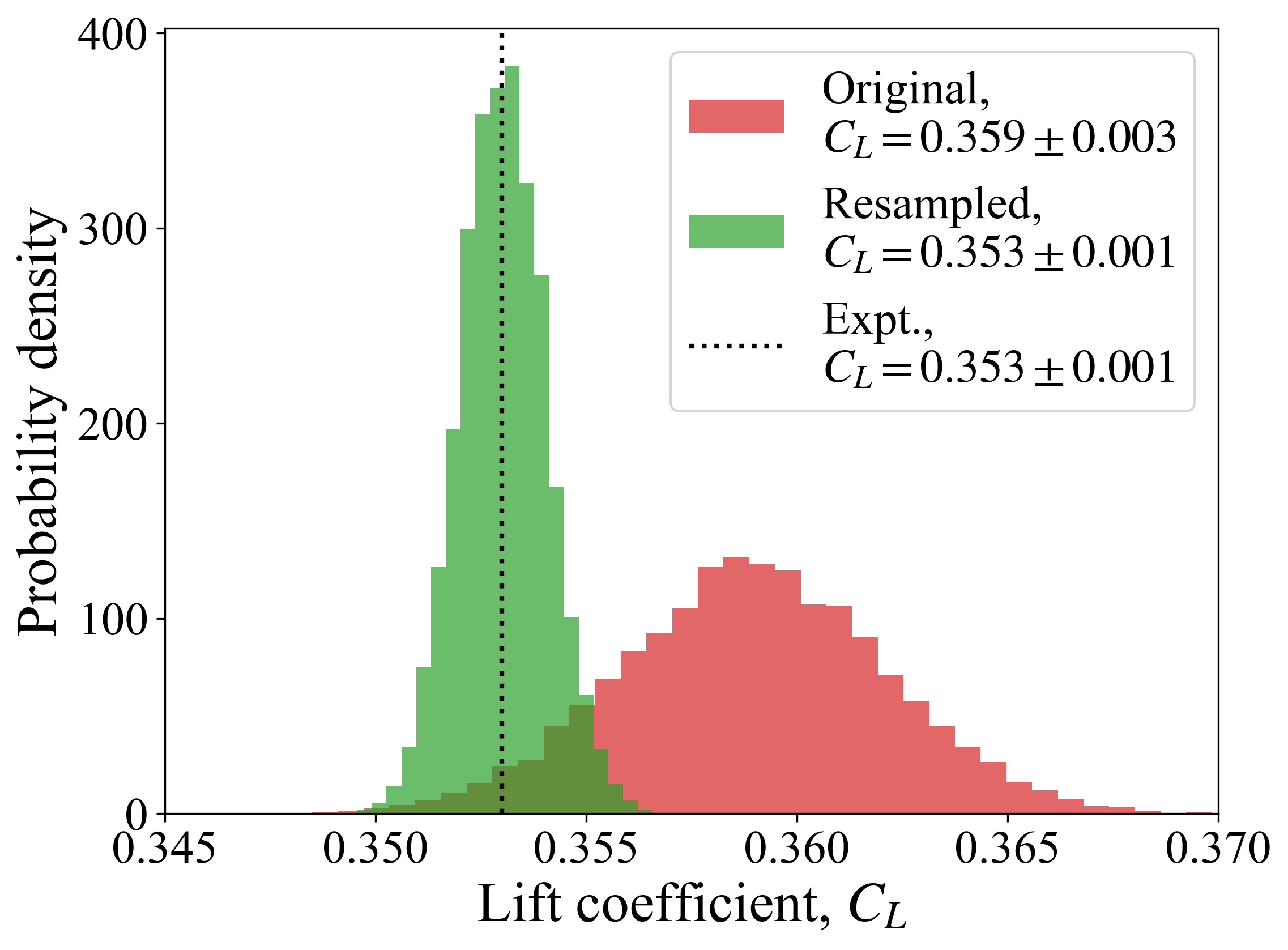}
\caption{Histogram of the samples from the original and the resampled predictive posterior distributions via conditioning on the known $C_L$. Presented results are for test case 3}
\label{fig:conditioning_hist}
\end{figure}

Figure~\ref{fig:conditioning_hist} provides a histogram-based visualization of the conditioning process. The samples from the initial unconditioned distribution (red) were more sparsely distributed and located farther from the true $C_L$. By selectively sampling from the predictive distribution using Eq.~\ref{eq:cond}, the conditioned samples (green) were restricted to a narrower range and at the prescribed mean. The figure illustrates that the newly drawn $C_p$ samples satisfy the condition that $C_L \sim \mathcal{N}(0.353 \pm 0.001)$.

\begin{figure}[h!]
\centering
\includegraphics[width=.85\textwidth]{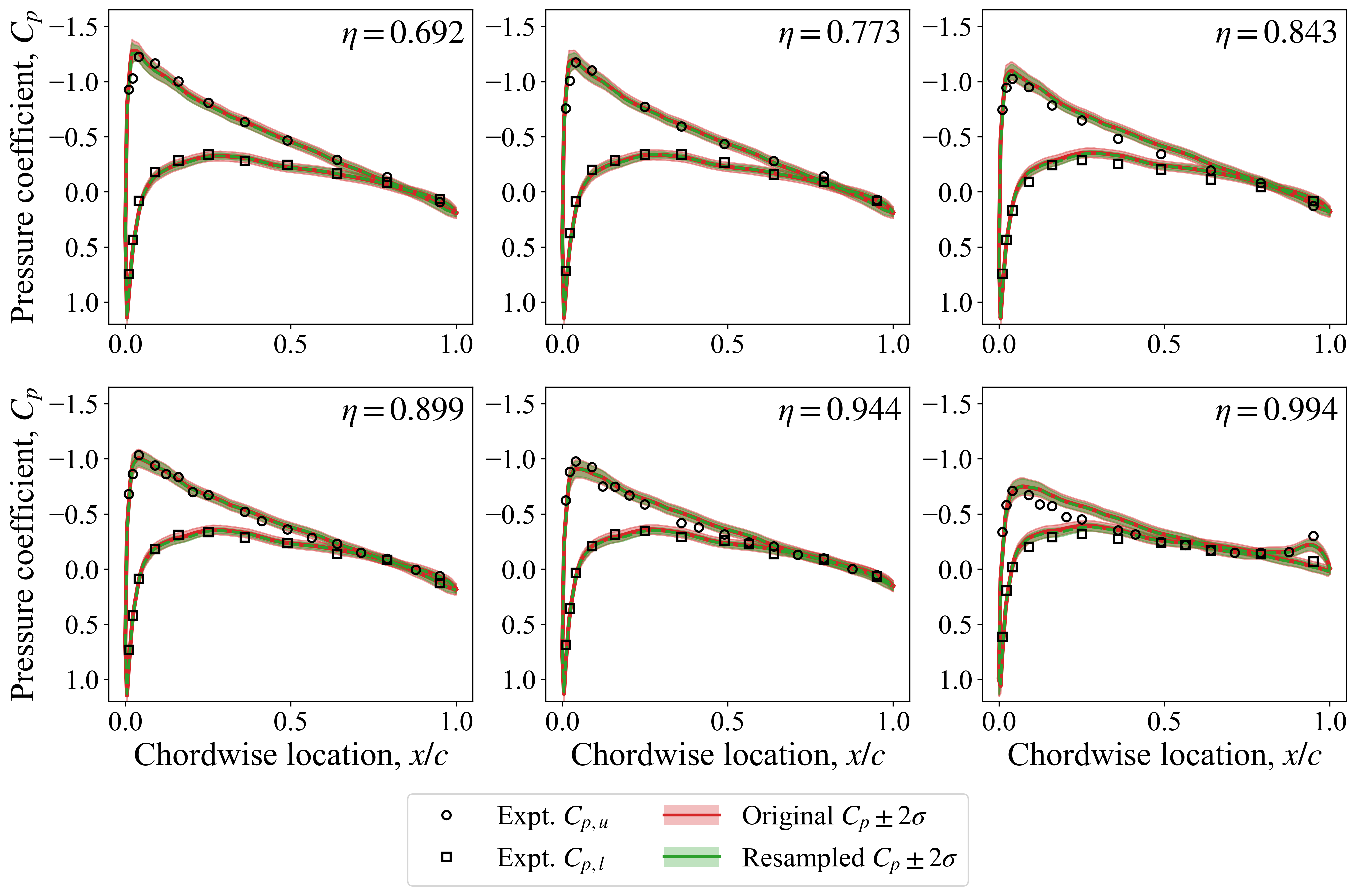}
\caption{Comparison between predicted chordwise $C_p$ distributions for model with and without $C_L$ conditioning. Presented results are for test case 3}
\label{fig:cond_slice}
\end{figure}

Lastly, Fig.~\ref{fig:cond_slice} illustrates how the $C_p$ predictions were influenced by posterior conditioning. Conditioning led to an increase in $C_p$ magnitude on the suction side and a decrease on the pressure side, resulting in improved agreement with the experimental measurements ($\text{MAE} = 0.052$ and $\text{MAE}_\text{enclosed} = 0.013$). Additionally, a reduction in posterior variance is observed when compared to predictions without conditioning. It is worth noting that these changes in $C_p$ are relatively small, underscoring the difficulty of accurately predicting $C_L$ from pressure distributions, an inherently sensitive task where even minor fluctuations in $C_p$ can have a significant impact.

\subsection{Computational Performance and Capabilities}
\begin{figure}[h!]
\centering
\includegraphics[width=.45\textwidth]{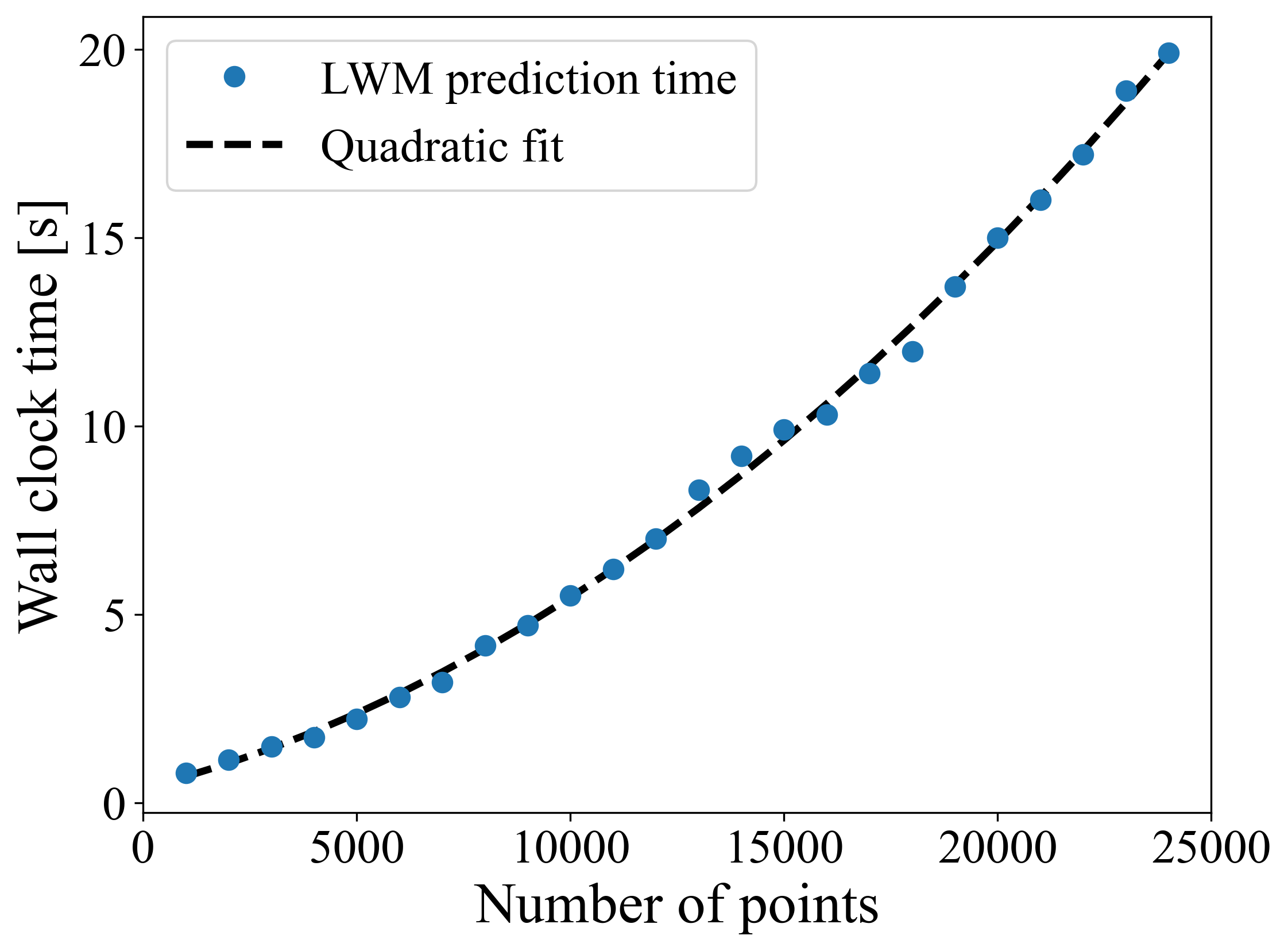}
\caption{Computational cost of the LWM as a function of the number of predicted points over a wing. The computations were performed on a single A100 GPU}
\label{fig:runtime_scalability}
\end{figure}

Considering that a sizable number of points is required to properly resolve the pressure distribution over an entire wing, it is valuable to understand how the model scales with the size of the test data. A Gaussian Process model is typically computationally limited by the need to invert an $N \times N$ covariance matrix, where $N$ represents the number of training data points. This inversion scales with a complexity of $\mathcal{O}(N^3)$. Since the inverted covariance matrix remains unchanged with respect to the test data, it can be pre-computed and stored in cache to bypass the computational burden. With this approach, the primary computational cost during prediction arises from computing the cross-covariance matrix for the LAM-prior and the predictive covariance (Eq.~\ref{eq:posteriorCov}), both of which scale with a complexity of $\mathcal{O}(M^2)$, where $M$ is the number of testing points. This quadratic scaling is evident in Fig.~\ref{fig:runtime_scalability}. Even at the highest number of points shown in the figure, the prediction takes 22.15 seconds on a single A100 GPU. This is a significant computational advantage compared to CFD simulations, which typically require tens of minutes to hours to complete depending on the complexity of the flow, not to mention the significant time and user knowledge required to generate the computational grid around the wing. 

As briefly discussed in Section~\ref{sec:intro}, most turbulence closures based on the Boussinesq hypothesis often fail to accurately capture wingtip vortex-induced effects in Reynolds-Averaged Navier–Stokes (RANS) solvers. Therefore, a comparison of capabilities between the Large Wing Model (LWM) and RANS was conducted, using experimental measurements as validation data. A RANS simulation was performed for test case 3, utilizing the mesh introduced in Section~\ref{subsec:gridflex}. Unlike the LWM, which utilizes only the 25,475 near-body points from the mesh, the CFD simulation requires the entire fluid domain, totaling 1.1 million points. The simulation was performed using ANSYS Fluent on an Intel(R) Xeon(R) E5-2660 v3 2.60 GHz CPU processor, parallelized across 30 local processes. In this simulation, the Spalart-Allmaras model was used as the turbulence closure.  

Instead of matching the experimental angle of attack used in test case 3, the angle was increased to 4.5 degrees to match the lift coefficient ($C_L$). The simulation was deemed converged when the residuals in lift and drag coefficients ($c_l$ and $c_d$) fell below $1 \times 10^{-5}$. The total runtime to reach convergence was approximately 29 minutes. In comparison, the LWM results were predicted on the near-body points with 54.11 seconds of wall clock time on the same CPU. 

\begin{figure}[h!]
     \centering
     \subfigure[Pressure coefficient contours of the LWM, experimental data, and RANS simulation. The horizontal black line indicates the spanwise station of $\eta = 0.994$.]{\includegraphics[width=0.75\textwidth]{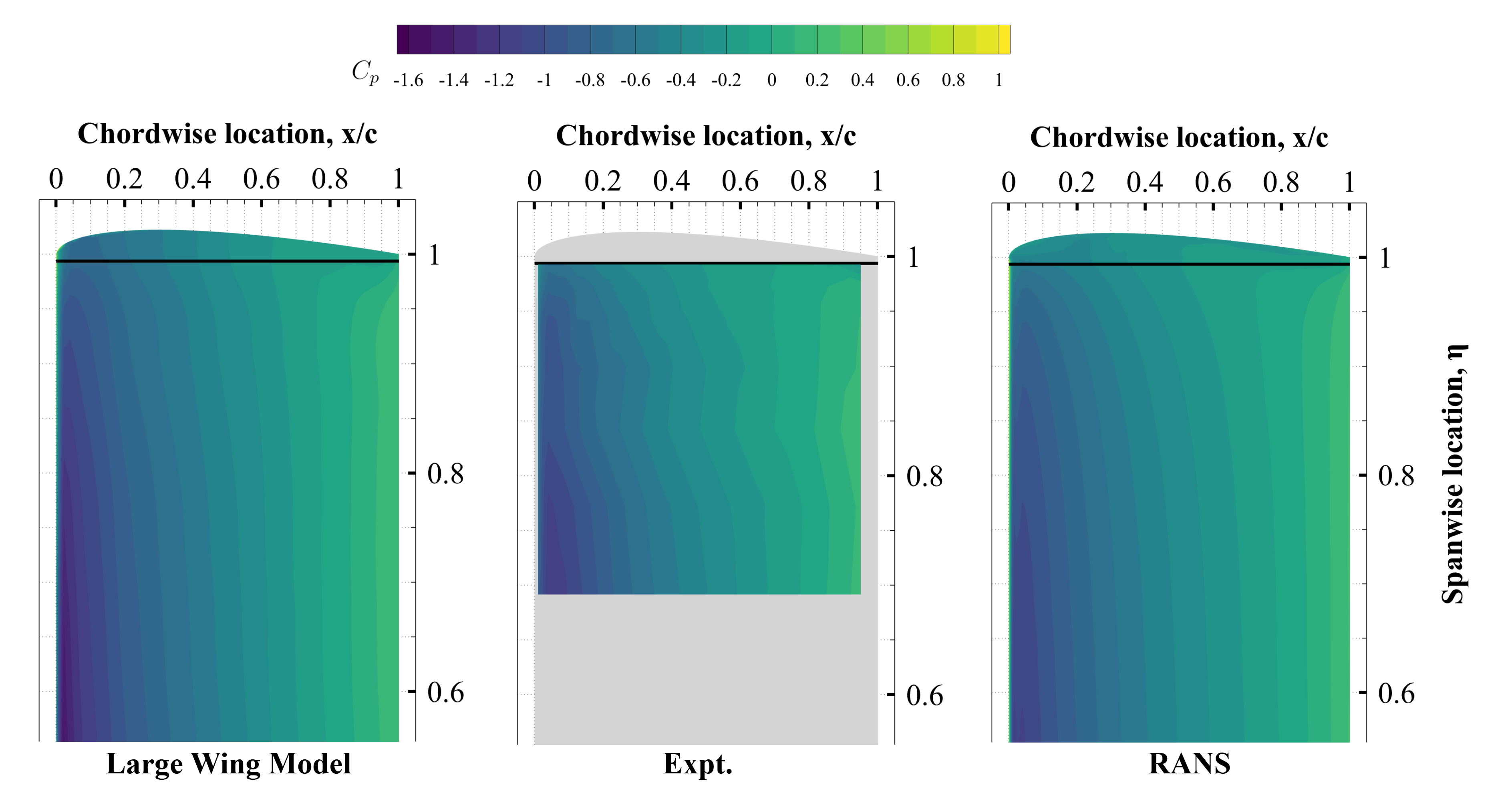}\label{fig:rans_contour}} 
     \subfigure[Chordwise slice of the wing $C_p$ distribution at the spanwise station of $\eta = 0.994$.]{\includegraphics[width=0.45\textwidth]{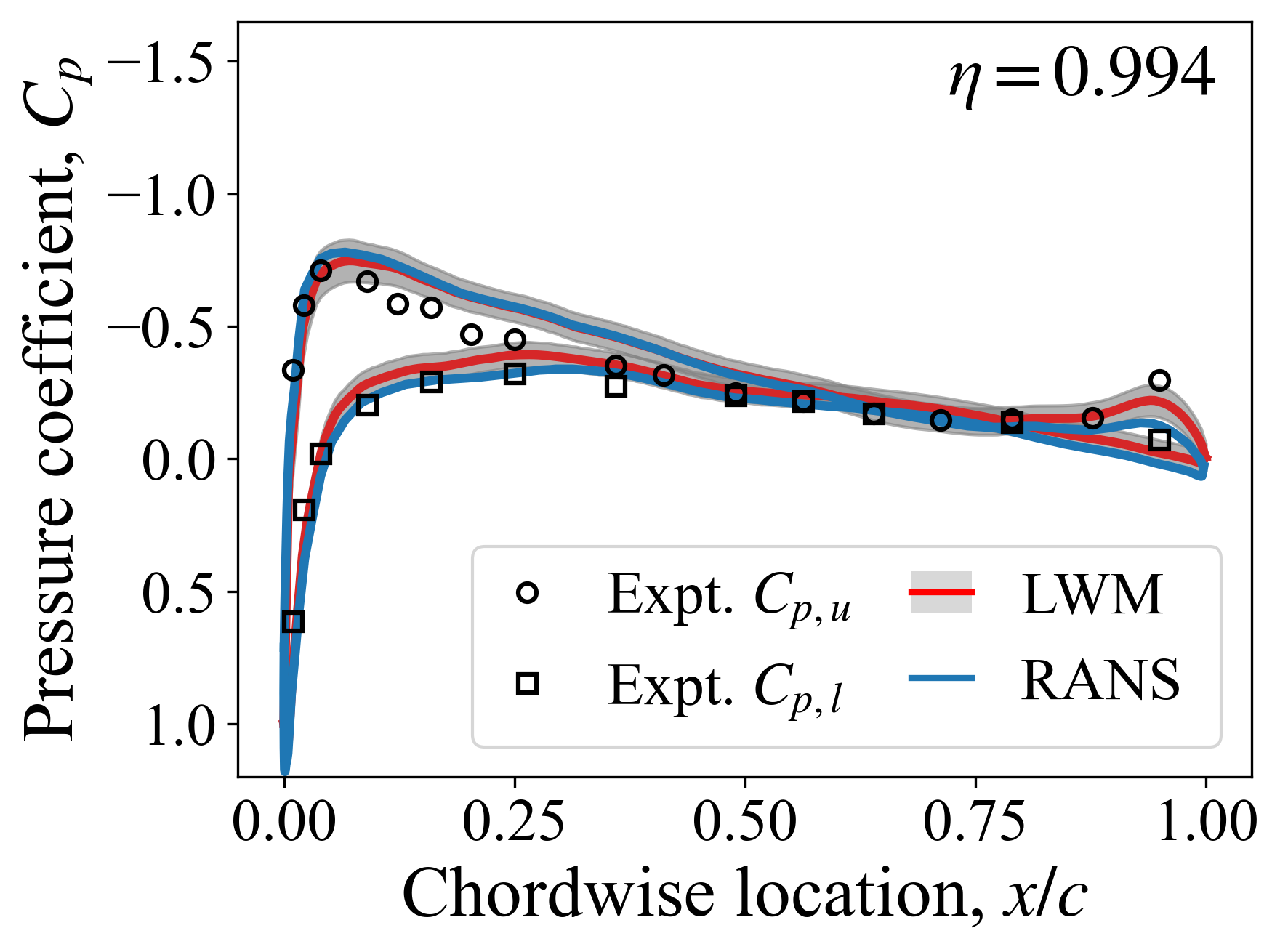}\label{fig:rans_slice}} 
     \caption{Comparisons between the LWM predictions, experimental data, and RANS simulation results. Presented results are for test case 3}
     \label{fig:rans_comparison}
\end{figure}

Figure~\ref{fig:rans_contour} provides a side-by-side comparison of the pressure contours generated by the LWM, experiment, and RANS simulation. The plots show that the mid-to-trailing edge pressures generally agree well across all three sources, until the very wing tip where pressure distortion becomes significant. A discrepancy is observed in the extent of this pressure distortion (upper right corner of the contour); the LWM predictions extend slightly more inboard compared to the experimental data. However, the presence and magnitude of the effect are reasonably captured by the model. In contrast, the RANS prediction severely underestimates both the magnitude of the $C_p$ and the spatial extent of the distortion in the wingtip trailing edge region. 

In addition, the LWM and RANS predictions exhibit discrepancies at the wing leading edge, most notably in the range $0.60 \le \eta \le 0.80$. While the suction peak predicted by the LWM tends to be sharper as shown in Fig.~\ref{fig:TC3_slice}, its $C_p$ magnitude aligns well with the experimental data. On the other hand, the RANS prediction shows a significant underestimation of the suction peak in this region compared to the experimental measurements.

Figure~\ref{fig:rans_slice} provides additional insight into how well the tip vortex-induced pressure distortion is captured. The trailing edge region clearly shows that the RANS prediction of $C_p$ (blue line) underestimates the pressure distortion compared to the LWM prediction (red line). In other parts of the wing section, the RANS results follow a similar trend to the LWM, reasonably capturing the suction peak, though they tend to overestimate $C_p$ magnitude in the region $0.10 \le x/c \le 0.35$. This is due to the fact that most turbulence closures based on the Boussinesq Hypothesis often fail to accurately capture wing tip vortex-induced effects when paired with RANS solvers~(Wilcox and Chambers,~\citeyear{wingtipvort_issue1};  Bradshaw,~\citeyear{wingtipvort_issue2}; Liu et al.,~\citeyear{LIU2016227}). While the underprediction could potentially be improved by refining the mesh or employing alternative numerical schemes, these adjustments often come with increased computational cost, added workload, and a higher level of required expertise.

\section{Conclusion}\label{sec:concl}
In this study, the Large Wing Model (LWM), a machine learning model derived from the Large Airfoil Model (LAM) designed to predict the aerodynamics of a three-dimensional (i.e. finite) wing, was presented. As a probabilistic machine learning model, the LWM has the capability to make uncertainty-aware predictions of wing pressure coefficient ($C_p$) distributions, sectional and total wing lift coefficients ($c_l$ and $C_L$). Its deep kernel learning architecture was tailored to handle the unstructured nature of a data set built from various experimental sources. 

To support the training of the LWM, a database of wind tunnel measurements of finite wings was created. The data were obtained from existing technical documents, which contain large quantities of embedded undigitized data points. This new database will be made publicly available as an addition to ASPIRE, a previously published, open-source database of airfoil surface pressure measurements. The digitized pressure measurements are available for wings of various airfoil sections, planform geometries, and operating conditions. The experimental noise from measurement and digitization was subsequently modeled and incorporated using Bayesian inference.

The LWM employs a fully connected neural network to transform the input variables into 14 latent variables. A Gaussian process model is then constructed within the space formed by the latent variables and the spanwise dimension. The separation of the spanwise dimension from the neural network allowed for a precise definition of the wing tip location, which was required for capturing finite wing effects. The kernel was constructed by an element-wise product of Mat\'ern 5/2 kernels for each latent dimension and a Mat\'ern 3/2 kernel with spatially varying hyperparameters in the spanwise dimension.

Given the relatively limited availability of wind tunnel measurements of a wing, developing a robust model from a sparse data set was crucial. This challenge was addressed by incorporating the LAM-predicted two-dimensional airfoil $C_p$ distribution as a physics-driven prior distribution. By encoding these fundamental airfoil characteristics into the model before training, the LWM achieved improved prediction accuracy and robustness.

The accuracy of the LWM was assessed for three test cases. The mean absolute error (MAE) in $C_p$, the mean absolute error in enclosed area ($\text{MAE}_\text{enclosed}$), and the absolute error in $C_L$ ($|\Delta C_L|$) were used as the metric. The accuracy was highest for test case 1, a NACA 0012 rectangular wing. The MAE in $C_p$ and $\text{MAE}_\text{enclosed}$ were found to be 0.044 and 0.014, respectively. $\Delta C_L$ was -0.008, corresponding to a -1.259\% error. For test case 2, a NACA 0012 wing with a different aspect ratio and a higher freestream Mach number, the MAE in $C_p$ and $\text{MAE}_\text{enclosed}$ were 0.062 and 0.029, respectively. Lastly, test case 3, a wing with an airfoil section geometry previously unseen by the model (NACA 0015), yielded a MAE of 0.052 in $C_p$ and 0.016 in enclosed area. The error in the lift coefficient, $\Delta C_L$, was 0.006, corresponding to a 1.7\% error. Notably, the tip vortex-induced pressure distortions were accurately resolved, a capability that lower-fidelity models (e.g. potential flow solvers) lack. 

In many scenarios, the integrated lift is either known or readily available, and the detailed pressure distribution ($C_p$) remains the primary quantity of interest. In such cases, the model posterior predictive space can be conditioned on the known $C_L$ information. An investigation of the effect of conditioning showed that the predicted $C_p$ values were adjusted to align with the given lift condition, resulting in improved prediction accuracy. For instance, by conditioning on the measured lift, the overall $\text{MAE}_\text{enclosed}$ for test case 3 was reduced from 0.016 to 0.013, a reduction of 18.75\%.

Lastly, the Gaussian Process model offers the ability to generate predictions at arbitrary spatial locations, providing enhanced flexibility over fixed-grid methods. This capability presents a distinct advantage: model outputs can be formatted to match existing CFD mesh structures, a widely adopted standard in the aerospace community. As a result, predictions from the LWM can be seamlessly integrated into common visualization tools, facilitating both accessibility and direct comparison with high-fidelity CFD solvers. This flexibility was leveraged in two key ways in this study. First, all model predictions were rendered as contour plots on volumetric wing representations using standard visualization software, as demonstrated in Section~\ref{sec:results}. Secondly, a direct side-by-side comparison of the LWM and RANS simulation was performed, which allowed for a clear evaluation of their respective capabilities in capturing key aerodynamic features. The model prediction required 22.15 seconds on an A100 GPU and 54.11 seconds on an Intel(R) Xeon(R) E5-2660 CPU, whereas the RANS simulation took 29 minutes on the same CPU. Thus, the LWM offers significantly lower computational cost while requiring minimal user expertise in flow modeling, a valuable advantage when rapid turnaround times are needed, such as during the design phase.

\begin{Backmatter}
\paragraph{Acknowledgments}\label{sec:ack}
The authors would like to thank Aaditya Hingoo, Tom Hoang, and Aanchal Save for their assistance in the digitization of the experimental data, and Zoelle Wong and Aaron Crawford for their assistance with the wing mesh creation for the RANS simulations. 

\paragraph{Funding Statement}
This research was partially funded by the U.S. Government under Cooperative Agreement No. W911W6-21-2-0001. The U.S. Government is authorized to reproduce and distribute reprints for Government purposes notwithstanding any copyright notation thereon.

The views and conclusions contained in this document are those of the authors and should not be interpreted as representing the official policies or position, either expressed or implied, of the U.S. Army Combat Capabilities Development Command (DEVCOM), Aviation \& Missile Center (AvMC), or the U.S. Government.

\paragraph{Competing Interests}
The authors declare no competing interests exist.

\paragraph{Data Availability Statement}
The data and code that support the findings of this study are fully open-source. They are available in  \url{https://github.com/hwlee924/Large-Airfoil-Model/} and \url{https://github.com/hwlee924/Large-Wing-Model/}.

\paragraph{Author Contributions}

\noindent Howon Lee: Data curation, Formal analysis, Investigation, Methodology, Software, Validation, Visualization, Writing/editing

\noindent Pranay Seshadri: Conceptualization, Investigation, Methodology, Project Administration, Supervision, Writing/editing

\noindent Juergen Rauleder: Conceptualization, Funding Acquisition, Investigation, Project Administration, Resources, Supervision, Writing/editing

\bibliographystyle{apalike}
\bibliography{reference}
\end{Backmatter}

\end{document}